\documentclass[12pt]{article}

\usepackage{geometry}
\usepackage{amsfonts,amsmath,amssymb}
\usepackage[normalem]{ulem}
\usepackage{cite}
\usepackage{subcaption}
\usepackage{cancel}
\usepackage{mathrsfs}
\usepackage{wasysym}
\usepackage{bbm,bm}
\usepackage{graphicx}
\usepackage[labelfont=small]{caption}
\usepackage{hyperref}
\usepackage[dvipsnames,svgnames,x11names]{xcolor}
\usepackage{wrapfig}
\usepackage[capitalize]{cleveref}
\usepackage{tocloft}

\newcommand\myshade{90}
\colorlet{mycitecolor}{JungleGreen}
\colorlet{mylinkcolor}{purple}
\colorlet{myurlcolor}{MediumSeaGreen}
\hypersetup{
    colorlinks = true,
    citecolor  = mycitecolor!\myshade!black,
    linkcolor  = mylinkcolor!\myshade!black,
    urlcolor   = myurlcolor!\myshade!black
}

\title{\vspace{-1.5cm}\textbf{Deterministic evolution of gauge fields through a singularity}}
\vspace{-12pt}
\author{{\large Martina Adamo,$^{(a)}$ Flavio Mercati $^{(b)}$} \\ \small  Departamento de F\'isica, Universidad de Burgos, 09001 Burgos, Spain.}

\date{\small\today}

\begin{document}

\maketitle

\let\svthefootnote\thefootnote
\let\thefootnote\relax\footnote{${(a)}$ \href{mailto:madamo@ubu.es}{madamo@ubu.es}, \href{mailto:marti.adamo@gmail.com}{marti.adamo@gmail.com}, ${(b)}$ \href{mailto:flavio.mercati@gmail.com}{flavio.mercati@gmail.com}}
\let\thefootnote\svthefootnote
\setcounter{footnote}{0}

\vspace{-1cm}

\begin{abstract}\noindent The nature of gravitational singularities has been questioned by some recent research, challenging the notion that classical determinism breaks down at these points. By allowing for dynamic changes in the orientation of spatial hypersurfaces, Einstein’s equations can be uniquely extended across singularities in certain symmetry-reduced models. A key step in this work was to reformulate the dynamical equations in terms of physical degrees of freedom. The singular behavior, it turns out, is confined to the gauge or unphysical degrees of freedom, and the physical ones evolve smoothly through the singularity. This paper builds on these findings, extending them to a model of gravity coupled with Abelian gauge fields in a homogeneous but anisotropic universe. The study reveals that near the big bang, the dynamics of geometry and gauge fields can be reformulated in a way that preserves determinism, provided there is a change of orientation at the singularity. Intriguingly, the gauge fields are shown to maintain their orientation through the singularity, unlike the spatial hypersurfaces. This suggests that the predicted orientation change of spatial hypersurfaces has physical significance, potentially allowing an observer to determine which side of the big bang they occupy. These results are proved to extend also to non-Abelian gauge fields with only one spatial component.
\end{abstract}

\section{Introduction}

One of the most remarkable predictions of classical general relativity (GR) is the existence of gravitational singularities. These are regions of the spacetime manifold where certain physical quantities become meaningless in a coordinate-independent way. In these regions, for example, some components of the stress-energy tensor may diverge, as well as some curvature invariants, or the geodesic equation may be singular (\textit{i.e.}, geodesic incompleteness, as predicted by the Penrose--Hawking singularity theorems \cite{PenroseSingularityTheorem,HawkingSingularityTheorem1,HawkingSingularityTheorem2,HawkingSingularityTheorem3,HawkingEllisBook}).

Currently, quantum gravity effects are considered the most promising approach to regularize gravitational singularities \cite{MisnerQuantumCosmologyI-1969,Berger1982,AHS,Bojowald_2015}, similar to how QED renders the energy of a point-like electric charge finite, thanks to the uncertainty principle \cite{Schwinger1948}. However, spacetime singularities differ significantly from those in electromagnetism. Unlike the latter, spacetime singularities arise directly from the evolution (via Einstein's equations) of regular initial data. This makes them physical predictions of the theory, while the diverging energy of a point-like charge is a consequence of the idealization of a point particle, which is introduced manually into the initial conditions.

One of the most remarkable implications of gravitational singularities is the apparent breakdown of determinism. In Lorentzian field theories like GR, classical determinism refers to the ability to uniquely predict the values of the physical fields anywhere within a region of spacetime known as the \textit{causal diamond}, provided that the initial values of these fields on some space-like region are given. The presence of a gravitational singularity appears to violate determinism, making it impossible to predict the values of the fields throughout a future causal cone originating at the singularity. The loss of predictability in GR around these regions can be summarized by Hawking's words: ``One does not know what will come out of a singularity'' \cite{HawkingPRD1976}.

Nevertheless, recent works \cite{ThroughTheBigBang,Sloan:2019wrz,Mercati_2019,Mercati_2022} have revealed the possibility of preserving determinism in certain symmetry-reduced models that exhibit big bang or black hole singularities. Ref.~\cite{ThroughTheBigBang} proved that, under a homogeneous but not necessarily isotropic \textit{ansatz}, it is possible to reformulate Einstein's equations in terms of a set of variables that satisfies a theorem of existence and uniqueness at the singularity. This result has been established in \cite{ThroughTheBigBang,Sloan:2019wrz,Mercati_2019,Mercati_2022} for the initial singularity of the \textit{Bianchi-IX model}, a homogeneous nonisotropic universe with an $S^3$ topology \cite{Misner_Mixmaster,MTW}, filled with \textit{stiff matter}. The presence of this type of matter source is necessary to regularize the eternal chaotic dynamics that would otherwise occur as the singularity is approached. In the absence of stiff matter, the Bianchi-IX singularity exhibits Misner's \textit{mixmaster} behavior \cite{Misner_Mixmaster,MTW}: as the singularity is approached, the spatial volume goes to zero, while the \textit{shape} degrees of freedom (which measure the anisotropy of the spatial metric) oscillate chaotically. This intricate motion persists indefinitely in coordinate time, with the shape variables oscillating an infinite number of times before reaching the singularity. However, the singularity itself is reached within a finite amount of proper time; the behavior is then that of an \textit{essential} singularity (analogous to $\lim_{x\to 0} \sin \left(1/x \right)$). This essential singularity prevents knowledge of the exact values of all the physical degrees of freedom at the singularity. The presence of stiff matter regularizes this behavior, ensuring that the system enters a final phase of \textit{quiescence}, \textit{i.e.}, a nonchaotic anisotropic collapse described by the \textit{Bianchi-I} (or \textit{Kasner}) model \cite{Barrow1978}. This condition is indispensable for extending the solution of Einstein's equation through the singularity since each physical degree of freedom must admit a well-defined limit at the big bang. 

It is possible to identify a set of physical variables that remain well defined at the singularity,\footnote{See \cite{Sloan:2018lim,Sloan:2020taf,Gryb:2021qix,Sloan:2021hwx,Sloan:2022exs} for a reformulation of the dynamics of GR as a non-Hamiltonian system based on these variables.} enabling the formulation of the equations of motion in a manner consistent with the \textit{Picard--Lindel\"of theorem} on the existence and uniqueness of solutions. This implies that the newly introduced regular variables continue to evolve uniquely through the singularity. As the singularity resides at the boundary of the configuration space, this suggests the need to extend the configuration space of GR. In \cite{ThroughTheBigBang}, this is achieved by allowing changes in the orientation of space. The interpretation of the regular variables beyond the singularity is as follows: they describe the geometry of spatial hypersurfaces with reversed orientations that lie beyond the big bang. Consequently, a ``second universe'' emerges from the singularity with an inverted spatial orientation.

In \cite{Mercati_2019}, it was conjectured that the preservation of determinism is not limited to homogeneous models such as Bianchi IX but is a general characteristic of realistic big bang and black hole singularities. First, quiescence, which was a key feature of the original result \cite{ThroughTheBigBang}, is not exclusive to models with a stiff matter source. As noted in \cite{Mercati_2019}, the Starobinksy model, which involves an effective action for gravity that includes the lowest-order quantum corrections to the Einstein--Hilbert action, also exhibits this characteristic (in addition to being the most promising candidate for explaining inflation \cite{Planck2018}). Therefore, it can be said that pure (semiclassical) gravity alone, without any matter sources, tends towards a quiescent behavior. 

Second, the assumption of homogeneity, which allowed the treatability of the models examined thus far, does not seem to be a prerequisite for the continuity results to hold. A strong evidence supporting this idea is provided by the Belinsky--Khalatnikov--Lifshitz (BKL) conjecture \cite{BKL}, which states that, as one approaches a space-like singularity, the time derivatives in Einstein's equations dominate over spatial derivatives, implying that the asymptotic dynamics is described by an (infinite) set of decoupled ordinary differential equations, one for each spatial point. These equations are identical to the equations of motion for a Bianchi-IX universe and, in quiescent models, they exhibit the continuation result under discussion. Interestingly, the BKL conjecture is essentially proven for universes with stiff matter sources \cite{AR}, providing strong indications that inhomogeneities will not change the result regarding continuation through singularities.

Furthermore, the models examined thus far have only included scalar matter fields, which, as we remarked, can be seen as an effective description of quantum-gravitational degrees of freedom, in the case of Starobinksy's model \cite{Mercati_2019}. It is commonly understood that ``matter does not matter'' near a singularity \cite{Matterdoesntmatter,Uggla2003,AHS}. In the vicinity of an isotropic Friedmann--Lemaître--Robertson--Walker solution, a simple scaling argument shows that the contributions to Friedmann's equations arising from Standard Model matter ($a^{-3}$), radiation ($a^{-4}$), the cosmological constant ($a^{0}$), and spatial curvature ($a^{-2}$) are all suppressed compared to the contribution of anisotropic shear, which scales as $a^{-6}$ (where $a$ denotes the FLRW scale factor). Here, by anisotropic shear, we refer to what we later term \textit{shape kinetic energy}, representing the term analogous to kinetic energy associated with the change in anisotropy parameters (visualize a scenario in which we are in close proximity to an initially isotropic spatial metric that is gradually losing its isotropy). It should be noted that the only exception to this behavior is scalar fields, which contribute to the Friedmann equations with terms that scale as $a^{-6}$. Nevertheless, this intuition does not hold when we delve deep into the anisotropic regime \cite{Sloan:2016efd}. If the pressure of matter sources becomes anisotropic, it can interact in a complex manner with the shape degrees of freedom, and it is not possible to demonstrate that matter or radiation decouples in the same way as in the isotropic regime. In an anisotropic universe, the motto ``matter does not matter'' does not hold. Therefore, the continuability of Einstein's equations through the big bang needs to be proven separately in the presence of matter or radiation fields.

In this paper, our focus is on radiation, specifically electromagnetic and Yang--Mills fields. We aim to provide a comprehensive analysis of their dynamics near a big bang singularity in a simplified model, namely, under the assumption of homogeneity. Our first objective is to rigorously prove that the radiation degrees of freedom truly decouple from the gravitational ones (in the sense that they disappear from the equations of motion of the latter), while being driven by their own evolution. This will be the first goal of this paper.

The second goal is to study how the gauge degrees of freedom evolve under the influence of the gravitational ones as we progress through the singularity. This question is intriguing because, although the orientation of spatial slices is reversed upon crossing the singularity, it is not evident whether this reversal can be observed, for example by means of parity-breaking tests like beta decay \cite{paritybeta}. It remains uncertain whether such tests could determine the side of the singularity we find ourselves on. For this, we need to know what happens to gauge fields and fermions, whether, for example, their direction is flipped or not. This paper takes the first step towards addressing this question by analyzing the behavior of the gauge fields.

Our results suggest that gauge fields do not reverse their direction across the singularity, although we cannot prove this yet in a fully general context. Our analysis is restricted to homogeneous gauge fields, and furthermore, it applies in full generality only to Abelian gauge groups. In the non-Abelian case, our analysis is limited to a ``one-dimensional'' \textit{ansatz}, meaning that both the gauge vector potential and its conjugate momentum are assumed to point in the same spatial direction throughout the evolution. The relaxation of these assumptions will be the focus of future investigations.

The paper is structured as follows: Section \ref{Sec2} provides a review of the Hamiltonian formulation of the Einstein--Maxwell--Klein--Gordon system. Sections \ref{Sec3} and \ref{Sec4} focus on the phase-space reduction to the homogeneous case, assuming a spatial topology of a three-sphere and the invariance under translations of both the metric and the gauge fields throughout the evolution. This \textit{ansatz} is compatible with the Hamiltonian evolution and reduces the degrees of freedom to a finite set, whose equations of motion are ordinary differential equations in time.

Section \ref{Sec5} introduces the Misner variables commonly used to discuss homogeneous universes with a three-sphere topology (Bianchi-IX models) and demonstrates the inevitability of the singularity even in the presence of gauge fields. In Section \ref{Sec6}, a further simplification is introduced through the one-dimensional \textit{ansatz} for the gauge fields discussed earlier. Under this \textit{ansatz}, the continuation result can be (relatively) easily proven.

Section \ref{Sec7} considers the relaxation of the one-dimensional \textit{ansatz} for Abelian gauge fields and the continuation result is proven. The extension of this result to $SU(2)$ and $SU(3)$ gauge fields under the one-dimensional \textit{ansatz} is detailed in \cref{SecB}. However, the relaxation of this \textit{ansatz} in non-Abelian gauge theories goes beyond the scope of the present paper. Finally, in Section \ref{Sec8}, we draw conclusions based on the knowledge gained thus far. In Table \ref{table} we summarize the notations used in the paper.
\begin{table}[!ht]
\bgroup
\def\arraystretch{1.5}
\begin{tabular}{p{0.6\linewidth}l}
\hline\hline
Greek indices are spacetime indices & $\mu , \nu , \rho ,$ \dots \,$\in \{0,1,2,3\}$
\\
\hline
Latin lowercase indices from the end of the alphabet are spatial indices & $i , j , k  ,$ \dots\, $\in \{1,2,3\}$
\\
\hline
Latin lowercase indices from the beginning of the alphabet are internal/dreibein indices & $a , b , c   ,$ \dots \, $\in \{1,2,3\}$
\\
\hline
Latin uppercase indices are internal $SO(N)$ indices & $I , J , K,$ \dots \,$\in \{1, 2, \dots, N^2-1\}$
\\
\hline\hline
\end{tabular}
\egroup
\captionsetup{width=.8\linewidth}
\caption{\footnotesize{Notations used in the paper.}}
\label{table}
\end{table}

\section[Hamiltonian formulation of Einstein--Maxwell--Klein--Gordon theory]{Hamiltonian formulation of Einstein--Maxwell--\\Klein--Gordon theory}
\label{Sec2}

Our goal is to extend the model of \cite{ThroughTheBigBang,Mercati_2019} to the case of the Einstein--Maxwell--Klein--Gordon system: GR minimally coupled with electromagnetism and a massless scalar field without potential, whose action is given by
\begin{equation} \label{action}
\begin{array}{c}
S = \displaystyle{\int} \! d^4 x \, \sqrt{- h} \left(  R - {\tfrac{1}{4}}  h^{\mu\nu}  h^{\rho\sigma} F_{\mu\rho} F_{\nu\sigma} -\tfrac{1}{2} h^{\mu\nu} \partial_\mu\Phi\,\partial_\nu \Phi \right) \,, ~~~
F_{\mu\nu} = \partial_\mu A_\nu - \partial_\nu A_\nu  \,.
\end{array}
\end{equation}
In the Arnowitt--Deser--Misner Hamiltonian formalism \cite{ADM}, the four-dimensional Lorentzian metric $h_{\mu\nu}$ (we use the mostly positive signature convention) is split into its spatial components $g_{ij}$, which serve as canonical variables, and four other fields: the \textit{lapse} scalar $N$ and the \textit{shift} vector $N^i$,\footnote{The spatial metric, shift, and lapse are a two-tensor, a vector, and a scalar field, respectively, under diffeomorphisms of the spatial hypersurface.} which are Lagrange multipliers because their time derivatives do not appear in the action. The relations between these quantities and the spacetime metric components are given by $h_{ij}=g_{ij}$, $h_{00} = -N^2+N^i N^j g_{ij}$, and $h_{0i} = h_{i0} = N_i$. We indicated the determinant of the four-dimensional metric with $h$.

These quantities have the following physical interpretations: the spatial components $g_{ij}$ represent the three-dimensional metric of equal-time hypersurfaces, the shift generates infinitesimal spatial translations along the hypersurfaces, and the lapse represents the proper time measured by observers moving orthogonally between neighboring hypersurfaces. The time derivatives of $g_{ij}$ are replaced, through a Legendre transform, by the conjugate momenta $\pi^{ij}=\left(\sqrt{g}/ {2N}\right)\left( g^{ik} g^{jl} - g^{ij} g^{kl} \right)\left( \dot{g}_{kl} - \pounds_{\vec{N}}\, g_{kl} \right)$, where $g^{ij}$ is the inverse matrix of $g_{ij}$, $g$ is the determinant of $g_{ij}$, and $\pounds_{\vec{N}}$ is the Lie derivative with respect to the shift $N^i$.

The Hamiltonian decomposition of the electromagnetic action is similar to the familiar one in Minkowski spacetime: the canonical variables are the spatial components of the electromagnetic potential $A_i$, while the time component $A_0$ (which is the scalar potential) acts as a fifth Lagrange multiplier. The spatial components of the Faraday tensor are given by $F_{ij} = \partial_i A_j - \partial_j A_i$. The momenta canonically conjugate to $A_i$ are the components of the electric field $E^i=\left(\sqrt{g}/ {N}\right) g^{ij} \left( F_{0j}-N^k F_{kj} \right)$.

Furthermore, the Klein--Gordon action contributes with a pair of canonical variables: the scalar field $\Phi$ and its canonically conjugate momentum, denoted as $\pi_\Phi$.

The following equal-time Poisson-bracket relations hold for conjugate pairs of variables:
\begin{equation}\label{GeneralPB}
\begin{aligned}
\{ g_{ij} (t,x) , \pi^{kl}(t,y) \} & = \tfrac{1}{2}\left(\delta^k_i\,\delta^l_j+\delta^l_i\,\delta^k_j\right) \delta^{(3)}(x-y) \,, \\
\{ A_i (t,x) ,E^j (t,y) \} & = \delta^j_i \, \delta^{(3)}(x-y) \,,\\
\{ \Phi (t,x) ,\pi_\Phi (t,y) \} & =  \delta^{(3)}(x-y)\,,
\end{aligned}
\end{equation}
while all other brackets are zero. The time evolution is governed by the total Hamiltonian, which, for our system, is a linear combination of the following constraints:
\begin{equation}\label{eq:EinsteinMaxwellConstraints}
\begin{aligned}
\mathcal{H}[N]  & = \int \! d^3 x \, N \biggl(  {\tfrac{1}{{\sqrt{g}}}}  \left( \pi^{ij} \pi_{ij} - \tfrac{1}{2} \pi^2 + {\tfrac{1}{2}} g_{ij} E^i E^j+\tfrac{1}{2}\pi_\Phi^2 \right) \\
&\qquad \qquad\qquad+ \sqrt{g} \left( {\tfrac{1}{4} } g^{ij}  g^{kl} F_{ik} F_{jl} - K +\tfrac{1}{2}g^{ij}\partial_i\Phi\,\partial_j\Phi \right)\biggr) \,,
\\
\mathcal{H}_i[N^i] & = \int \! d^3 x \, N^i \left( E^j F_{ij} - 2 \, g_{ij} \nabla_{\!k} \pi^{jk} +\pi_\Phi\partial_i\Phi\right)  \,,
\\
\mathcal{G}[A_0] & = - \int \! d^3 x \, A_0 \left( \nabla_{\!i} E^i \right)
\,.
\end{aligned}
\end{equation}
Here, $K$ is the Ricci scalar and $\nabla_{\!i}$ is the covariant derivative, both with respect to the metric $g_{ij}$. These constraints are first class, and close an extension of the so-called \textit{hypersurface deformation algebra} \cite{HOJMAN197688,FlavioSDbook} (or rather \textit{algebroid} \cite{Blohmann:2010jd}). The first and last lines in \cref{eq:EinsteinMaxwellConstraints} represent the Hamiltonian and Gauss constraints, responsible for generating time evolution and electromagnetic gauge transformations, respectively. The three constraints $\mathcal{H}_i$ can be expressed (up to boundary terms) as a linear combination of Gauss and diffeomorphism constraints:
\begin{equation}
    \mathcal{H}_i[N^i]  = \mathcal{D}_i[N^i] - \mathcal{G}[ A_i  N^i] + \text{(boundary terms)} \,,
\end{equation}
where the diffeomorphism constraints are defined as 
\begin{equation}
\mathcal{D}_i[N^i]  = \int \! d^3 x \left(   E^i \pounds_{\vec N} A_i  +  \pi^{ij} \pounds_{\vec N} g_{ij} +\pi_\Phi \pounds_{\vec N} \Phi \right)   \,,
\end{equation}
and generates spatial diffeomorphisms.

\section{Homogeneous \textit{ansatz}}
\label{Sec3}

We now impose the homogeneous \textit{ansatz}, which has been the starting point of previous works such as \cite{ThroughTheBigBang,Mercati_2019}.  This assumption can be motivated with the aforementioned BKL phenomenon \cite{BKL}, which implies that inhomogeneities are suppressed near the singularity, and decouple from the equations of motion of the homogeneous degrees of freedom. To describe the dynamics of the universe near the singularity, therefore, one needs to begin by considering the homogeneous degrees of freedom. The homogeneous cosmological models are based on Bianchi's classification of homogeneous three-dimensional geometries \cite{Bianchi1898,Misner_Mixmaster,MTW}. Among these, the simplest model with a nontrivial dynamics is ``Bianchi IX,'' in which the spatial topology is that of a three-sphere $S^3$ \cite{ThroughTheBigBang,Mercati_2019}. Technically, homogeneity means that the spatial metric is assumed to have three independent Killing vectors that generate spatial translations. On $S^3$, coordinatized by the usual hyperspherical coordinates $\theta\in [0,\pi]$, $\phi \in [0,\pi]$, $\psi\in [0, 2\pi)$, it is possible to construct a basis of vector fields that are invariant under these translations, as well as a dual basis of one-forms:
\begin{align}
&\left\{
\begin{aligned}
&\chi_1 =  (4\pi^2)^{\frac 1 3} \left( -\sin \psi \,  \partial_\theta + \cos \psi \, \csc \theta \, \partial_\phi  -  \cos \psi \, \cot \theta \,\partial_\psi 
\right)\,, \\
&\chi_2 =  (4\pi^2)^{\frac 1 3} \left( - \cos \psi \,  \partial_\theta - \sin \psi \, \csc \theta \,  \partial_\phi +  \sin \psi \, \cot \theta \, \partial_\psi \right)\,,
\\
&\chi_3 =  (4\pi^2)^{\frac 1 3} \, \partial_\psi \,,
\end{aligned}
\right. 
\\ \nonumber \\ 
&\left\{
\begin{aligned}
&\sigma^1 =  (4\pi^2)^{-\frac 1 3} \left( -\sin \psi \, d \theta + \cos \psi \, \sin \theta \, d \phi \right)\,,
\\
&\sigma^2 =  (4\pi^2)^{-\frac 1 3} \left(  -\cos \psi \, d \theta - \sin \psi \, \sin \theta \, d \phi \right)\,,
\\
&\sigma^3 =   (4\pi^2)^{-\frac 1 3} \left( d \psi + \cos \theta \, d \phi \right)\,,
\end{aligned}
\right.
\end{align}
where the normalization factors are chosen so that the integral of the volume form is one, $\int \!d^3x\, \det \sigma = \int_0^{2\pi}\!\int_0^\pi\!\int_0^\pi\!d\theta d\phi d\psi\, \sin \theta  = 1$. The duality between $\chi_a^i$ and $\sigma^a_i$ is
\begin{equation}\label{DualityVectors1forms}
\sigma^a_i \chi_a^j = \delta^j_i \,, \qquad 
\sigma^a_i \chi_b^i = \delta^a_b \,. 
\end{equation}
The most generic homogeneous (but not necessarily isotropic) metric on $S^3$ can be expressed as a quadratic form (with spatially constant coefficients) in this basis. By also imposing the homogeneous \textit{ansatz} on the conjugate momenta $\pi^{ij}$ (which is necessary to preserve the homogeneous \textit{ansatz} for the metric under time evolution), as well as on the electromagnetic fields $A_i$ and $E^i$, and on the scalar fields $\Phi$ and $\pi_\Phi$, we obtain
\begin{equation}
\begin{array}{l l}
\begin{aligned}
g_{ij} (t,x) & = q_{ab}(t) \, \sigma^a_i(x) \, \sigma^b_j(x) \,, \\
A_i (t,x) & = A_a (t)\, \sigma^a_i(x) \,,\\
\Phi(t,x) &=q_0(t)\,,
\end{aligned}
\begin{aligned}
\quad \pi^{ij} (t,x) & = p^{ab}(t)  \,  \chi^i_a(x) \, \chi^j_b(x) \, \det \sigma(x) \,,
\\
E^i (t,x) & = E^a(t) \, \chi^i_a(x) \, \det \sigma(x) \,, \\
\pi_\Phi(t,x)&=p^0(t) \, \det \sigma(x)\,,
\end{aligned}
\end{array}
\end{equation}
Here, $t$ and $x=(\theta,\phi,\psi)$ denote the dependence on time and spatial (hyperspherical) coordinates, respectively. Due to the homogeneous \textit{ansatz}, the Faraday tensor and the determinant of the metric in the invariant basis become
\begin{equation}
\begin{aligned}
        F_{ij} (t,x) &= F_{bc} (t) \,\sigma^b_i (x) \, \sigma^c_j (x)  =  -A_a (t) \, \delta^{ad} \, \varepsilon_{dbc} \, \sigma^b_i (x) \, \sigma^c_j (x) \,, 
    \\
    g(t,x)&=q(t) \, (\det \sigma(x))^2\,,
\end{aligned}
\end{equation}
where $q$ denotes the determinant of $q_{ab}$. Notice that the conjugate momenta $p^{ij}$, $E^i$, and $\pi_\Phi$ require the term $\det \sigma (x)=\sin \theta$ to ensure the correct transformation behavior under diffeomorphisms, which is that of a tensor density with a weight of +1. In this basis, all these tensor fields have homogeneous components: $q_{ab}$, $p^{ab}$, $A_a$, $E^a$, $q_0$, $p^0$, with only a time dependence. 
We can deduce the Poisson brackets between the homogeneous components from \cref{GeneralPB} and \cref{DualityVectors1forms}. For example, in the case of the scalar field, one has that
\begin{equation}
    q_0 (t) = \int \! d^3 x \,\Phi(t,x)  \det  \sigma(x)  \,, \qquad
    p^0 (t) = \int\!d^3 x\, \pi_\Phi(t,x)  \,.
\end{equation}
Hence, their Poisson bracket is
\begin{equation}
\begin{aligned}
        \{ q_0 , p^0 \} &= \int\! d^3 x\, d^3 y \, \det \sigma(x) \{\Phi(t,x) , \pi_\Phi(t,y)\} = \int\! d^3 x\, d^3 y\, \det \sigma(x) \, \delta^{(3)}(x-y) \\
        &= \int d^3 x \, \det \sigma (x) = 1 \,.
\end{aligned}
\end{equation}
The same procedure works similarly for the gauge and metric fields (whose indices, however, need to be saturated with an appropriate number of  basis vector and basis one-form fields $\chi^i_a$, $\sigma_i^a$), and one ends up with the following Poisson brackets:
\begin{equation}
\{ q_0, p^0 \}  = 1\,, \qquad
\{q_{ab} , p^{cd} \}  = \tfrac{1}{2}\left(\delta^c_a\,\delta^d_b+\delta^d_a\,\delta^c_b\right)  \,, \qquad
\{ A_a ,E^b  \}  = \delta^b_a \,.
\end{equation}
Under the homogeneous \textit{ansatz}, all the constraints (each of which constrains one degree of freedom per spatial point) can be smeared over arbitrary functions and turn into the following global constraints:
\begin{equation} \label{constraints}
\begin{aligned}
\mathcal{H}[N] & = n \left(  p^{ab}p^{cd}q_{bc}\,q_{da} - \tfrac{1}{2} (p^{ab}q_{ab})^2 + q_{ab}\,q_{cd}\,\delta^{bc}\delta^{da}  - \tfrac{1}{2} ( q_{ab}\,\delta^{ab} )^2   \right.
\\ 
& \qquad \qquad\qquad \qquad\qquad \left. +\tfrac{1}{2}(p^0)^2+ \tfrac{1}{2} q_{ab}\, E^a E^b  +   \tfrac{1}{4} q \, q^{ab}  q^{cd} F_{ac}\, F_{bd}  \right) \,,
\\ \\
\mathcal{H}_i[N^i]  &= n^d \left( E^a F_{da}   + 2 \, p^{ab} \,q_{ac} \, \varepsilon_{bdf}\,\delta^{fc}  \right) = \mathcal{D}_i [N^i] \,,
\\ \\
\mathcal{G}[\varphi]  & =0\,,
\end{aligned}
\end{equation}
where
\begin{equation}\label{Lagrangemult}
n = \tfrac{1}{\sqrt{ q}}\int \! d \theta\, d\phi\, d \psi  \,  \sin \theta \, N(x) \,,
\qquad
n^a  = \int\! d \theta\, d\phi\, d \psi \,  \sin \theta \,  \sigma^a_i(x) N^i(x) \,,
\end{equation}
are four leftover Lagrange multipliers (the spatial averages of the lapse and shift).
Notice that the Gauss constraint is automatically solved, and that the scalar field only contributes with a kinetic term in the global Hamiltonian constraint. The homogeneous \textit{ansatz} is dynamically consistent, \textit{i.e.}, it is preserved by the evolution \cite{MTW,Misner_Mixmaster}, as can be verified by substituting it into the right-hand side of the Einstein equations.

\section{Solving the diffeomorphism constraints}\label{DiffeoSolution}
\label{Sec4}

In order to eliminate the nonphysical degrees of freedom, we need to gauge fix the three diffeomorphism constraints using Dirac's procedure for constrained Hamiltonian systems \cite{DiracLecture,Teitelboim,HansonTeitelboim}. The diffeomorphism generators appearing in \cref{Lagrangemult} are $\xi_a = E^a F_{da}   + 2 \, p^{ab} \,q_{ac} \, \varepsilon_{bdf}\,\delta^{fc} \approx 0$, that is
\begin{equation} 
	\left\{\begin{aligned}
		\xi_1  & = 2 \left( p^{13} q_{12} - p^{12} q_{13} + p^{23} q_{22} - p^{22} q_{23} + p^{33} q_{23} - p^{23} q_{33} \right) + E^3 A_2 - E^2 A_3\,,  \\
		\xi_2 & = 2 \left( p^{11} q_{13} - p^{13} q_{11} + p^{12} q_{23} - p^{23} q_{12} + p^{13} q_{33} - p^{33} q_{13} \right) + E^1 A_3 - E^3 A_1\,,  \\
		\xi_3 & = 2 \left( p^{22} q_{12} - p^{12} q_{22} + p^{23} q_{13} - p^{13} q_{22} + p^{12} q_{11} - p^{11} q_{12} \right) + E^2 A_1 - E^1 A_2 \,.
	\end{aligned}\right.
\end{equation}
A suitable choice \cite{FlavioSDbook,Mercati_2019} for the gauge-fixing constraints is
\begin{equation} \label{gaugefixing}
\xi_4 = q_{23} \approx 0 \,,
~~
\xi_5 =  q_{13} \approx 0 \,,
~~ 
\xi_6  = q_{12} \approx 0 \,.
\end{equation}
These gauge-fixing constraints are second class with respect to $\xi_1$, $\xi_2$, $\xi_3$ everywhere except for the three planes of symmetry $q_{11} = q_{22}$, $q_{22} = q_{33}$, and $q_{33} = q_{11}$. Apart from a measure-zero set of solutions that lives entirely on these planes, all solutions intersecting these planes can be uniquely continued through them by making a different local choice of gauge fixing. We can solve the diffeomorphism constraints with respect to the nondiagonal components of $p^{ab}$, and this choice is regular everywhere away from the three symmetry planes
\begin{equation}
    p^{23} = \frac  { E^2  A_3 - E^3  A_2  }{2  \left( q_{22} - q_{33}  \right)}\,, \qquad
    p^{13}  = \frac  { E^3  A_1 - E^1  A_3}{2  \left( q_{33} - q_{11}  \right)}\,, \qquad
    p^{12}  = \frac  { E^1  A_2 - E^2  A_1 }{2  \left( q_{11} - q_{22}  \right)}\,.
\end{equation}
Using the six second-class constraints $\xi_\alpha$, $\alpha=1,\dots,6$, we can construct the Dirac matrix, which, when evaluated on the constraints hypersurface in the space of solutions of the system (\textit{i.e.}, on shell), reads
\begin{equation}
C_{\alpha\beta} =
\left\{ \xi_\alpha , \xi_\beta \right\}  \approx 
\left(
\begin{array}{c|c}
0 & M
\\
\hline
-M   & 0
\end{array}
\right) \,,
\end{equation}
where $M=\operatorname{diag}(q_{33}-q_{22},q_{11}-q_{33},q_{22}-q_{11})$. The inverse Dirac matrix is then simply
\begin{equation}
(C^{-1})^{\alpha\beta}   \approx 
\left(
\begin{array}{c|c}
0 & M^{-1}
\\
\hline
 -M^{-1}   & 0
\end{array}
\right) \,.
\end{equation}
Therefore, the Dirac bracket
\begin{equation}
\{ f , g\}_* = \{ f , g\} - \{ f , \xi_\alpha \}(C^{-1})^{\alpha\beta}    \{ \xi_\beta , g\} \,,
\end{equation}
is canonical on the diagonal components of the metric and their momenta, the three components of the electromagnetic potential and their momenta, and the scalar field and its momentum. At this point, it is convenient to simplify the notation for the diagonal components of the metric and momenta:
\begin{equation}
    q_1 = q_{11} \,, \quad q_2 = q_{22} \,, \quad q_3= q_{33} \,, \qquad  p^1 = p^{11} \,, \quad p^2 = p^{22} \,, \quad p^3= p^{33} \,.
\end{equation}
In these variables, the Dirac brackets read
\begin{equation}
\{ q_0, p^0 \}_*  = 1\,, \qquad \{ q_a , p^b \}_* = \delta^b_a \,, \qquad \{A_a , E^b \}_* = \delta^b_a \,,
\end{equation}
and all the other Dirac brackets are zero.

We began with a system described by 20 degrees of freedom (although not all of them are physical): six components of the symmetric three-dimensional metric $q_{ab}$, six components of metric momenta $p^{ab}$, three components of the electromagnetic potential $A_a$, three of the electromagnetic momenta $E_a$ (representing the electric field), one component of the scalar field $q_0$, and one of the scalar momentum $p^0$. After applying the diffeomorphism gauge fixing, we constrain six degrees of freedom. Specifically, we set to zero the three off-diagonal components of the metric, $q_{12}$, $q_{23}$, $q_{13}$, and transform the three off-diagonal components of the metric momenta, $p^{12}$, $p^{23}$, $p^{13}$, into functions of all the other variables. As a result, we were left with 14 degrees of freedom. Among these, 12 are genuinely physical, meaning they are the minimum number of independent variables required to uniquely determine a solution. The remaining two degrees of freedom are subject to constraints imposed by the Hamiltonian constraint (the first equation in \eqref{constraints}) and its gauge-fixing condition, which allows us to freely choose initial conditions along the solution curve without altering the solution itself.

The equations of motion generated by the Dirac bracket are the canonical equations of motion obtained from the on-shell Hamiltonian
\begin{equation}\label{GeneralBIX+GaugeHamiltonianConstraint}
\begin{aligned}
\mathcal{H}[N]  &=n \, \bigg{(} \mathcal{H}_{BIX} +\tfrac{1}{2}(p^0)^2+ \frac{q_2 q_3 (M_1)^2}{2\left(q_2-q_3\right)^2}+\frac{q_1 q_3 (M_2)^2}{2\left(q_1-q_3\right) ^2}+\frac{q_1 q_2 (M_3)^2}{2\left(q_1-q_2\right)^2}
\\ 
& \qquad ~ + \tfrac{1}{2} q_1 \left((E^1)^2 + (A_1)^2 \right)  + \tfrac{1}{2} q_2 \left((E^2)^2 + (A_2)^2 \right) + \tfrac{1}{2} q_3 \left((E^3)^2 + (A_3)^2 \right)
  \bigg{)} \,,
\end{aligned}
\end{equation}
where we defined
\begin{equation}\label{DefinizioneVettoreM}
M_1  = E^2  A_3 - E^3  A_2 \,,
~~~
M_2  =   E^3  A_1 - E^1  A_3 \,,
~~~
M_3=  E^1  A_2 - E^2  A_1 \,, 
\end{equation}
and
\begin{equation}\label{EmptyBIXham}
\begin{aligned}
    \mathcal{H}_{BIX} =& \left(p^1 q_1\right)^2+\left(p^2 q_2\right)^2+\left(p^3 q_3\right)^2 - {\tfrac 1 2} \left(p^3 q_3+p^2 q_2+p^1 q_1\right)^2 \\
    &+ q_1^2+q_2^2+q_3^2-{\tfrac 1 2} \left(q_1+q_2+q_3\right)^2  \,,
\end{aligned}\end{equation}
is the Hamiltonian constraint of an empty Bianchi-IX universe \cite{Mercati_2019}.

\section{Inevitability of collapse}
\label{Sec5}

Consider the following canonical transformation:
\begin{equation}\label{CanonicalTransfMisnerVariables}
\left\{
\begin{aligned}
&q_0= x^3 \,,
\\
&q_1=a_0^2 \, \exp \left(\tfrac{x^0-\sqrt{3} x^1+x^2}{\sqrt{3}}\right) \,,
\\
&q_2=a_0^2 \,\exp \left(\tfrac{x^0+\sqrt{3} x^1+x^2}{\sqrt{3}}\right)\,,
\\
&q_3=a_0^2 \,\exp \left(\tfrac{x^0-2 x^2}{\sqrt{3}}\right)\,,
\end{aligned}
\right.
\qquad
\left\{
\begin{aligned}
&p^0= k_3 \,,
\\
&p^1=a_0^{-2} \,\left(\tfrac{ k_2-\sqrt{3} k_1+2 k_0}{2 \sqrt{3}}
\right) \exp \left(-\tfrac{x^0-\sqrt{3} x^1+x^2}{\sqrt{3}}\right) \,,
\\
&p^2=a_0^{-2}\left(\tfrac{ k_2+\sqrt{3} k_1+2 k_0}{2 \sqrt{3}} \right) \exp \left(-\tfrac{x^0+\sqrt{3} x^1+x^2}{\sqrt{3}}\right)\,,
\\
&p^3=a_0^{-2}\left(\tfrac{ k_0-k_2}{\sqrt{3}} \right) \exp \left(-\tfrac{x^0-2 x^2}{\sqrt{3}}\right)\,,
\end{aligned}
\right.
\end{equation}
where $a_0$ is a dimensional constant (a reference scale). Since the transformation is canonical, the new variables are canonically conjugate to each other:
\begin{equation}
\{ x^0 , k_0 \}_* =1 \,, \qquad 
\{ x^a , k_b \}_* = \delta^a_b \,.     
\end{equation}
In the new variables, the Hamiltonian constraint takes the form of a diagonal quadratic kinetic term for the metric variables $k_0,k_1,k_2,$ and the scalar variable $k_3$, along with a potential-like term dependent on the metric and electromagnetic variables:
\begin{equation}\label{H3d}
\begin{array}{c}
\mathcal{H}[N] = n \left( \tfrac{1}{2} \left( - k_0^2 +  k_1^2  +  k_2^2+k_3^2 \right)  + {\tfrac{1}{2}} U (x,A,E) \right) \,, \\ \\
U(x,A,E) =  a_0^4 \, e^{\tfrac{2 x^0}{\sqrt{3}}} C(x) +  a_0^2\, e^{\frac{x^0}{\sqrt{3}}} V(x,A,E) +  W(x,A,E)  \,.
\end{array}
\end{equation}
In the previous equation,
\begin{equation}\label{BIXpot}
\begin{aligned}
C(x) & =  e^{-2x^1 + \tfrac{2}{\sqrt{3}} x^2} + e^{2x^1 + \tfrac{2}{\sqrt{3}} x^2} + e^{-\tfrac{4}{\sqrt{3}} x^2}   -2 \left( e^{\tfrac{2}{\sqrt{3}} x^2} + e^{-x^1 - \tfrac{1}{\sqrt{3}} x^2} + e^{x^1 - \tfrac{1}{\sqrt{3}} x^2}\right) \, ,
\end{aligned}
\end{equation}
is the Bianchi-IX potential \cite{Mercati_2019,FlavioSDbook}, and
\begin{equation}
	\begin{aligned}\label{VandWpots}
		&V(x,A,E)  =  e^{-x^1+\tfrac{x^2}{\sqrt{3}}} \left( (E^1)^2 + A_1^2 \right)  + e^{x^1+\frac{x^2}{\sqrt{3}}} \left( (E^2)^2 + A_2^2 \right) + e^{-\frac{2x^2}{\sqrt{3}}} \left( (E^3)^2 + A_3^2 \right) , \\ \\
		&W(x,A,E)  = \frac{e^{x^1+\sqrt{3}x^2}(M_1)^2}{(e^{x^1+\sqrt{3}x^2}-1)^2} + \frac{e^{x^1+\sqrt{3}x^2}(M_2)^2}{(e^{x^1}-e^{\sqrt{3}x^2})^2} + \frac{e^{2x^1}(M_3)^2}{(e^{2x^1}-1)^2} \, ,
	\end{aligned}
\end{equation}
are two contributions depending on the electromagnetic field.

The variables $x^1$ and $x^2$ represent the \textit{shape} degrees of freedom, which quantify the anisotropy of the spatial metric. Their conjugate momenta, $k_1$ and $k_2$, correspond to the rate of change of these anisotropies. The variable $x^0$ and its conjugate momentum $k_0$ are related to the volume of the universe $v$ and its conjugate momentum $\tau$, known as the \textit{York time}. These relationships are expressed as follows:
\begin{equation}\label{def of volume}
	v = a_0^3 \, e^{\frac{\sqrt{3}}{2} x^0} \, , \qquad \tau = \tfrac{2}{\sqrt{3}} \, a_0^{-3}\, e^{-\frac{\sqrt{3}}{2} x^0} k_0 \, .
\end{equation}
The variable $\tau$, named after James York \cite{Yorktime}, is associated with a specific foliation of spacetime known as the \textit{constant-mean extrinsic curvature}. In this foliation, the initial-value problem can be formulated as a system of elliptic equations, whose solution exists and is unique. In a cosmological setting, $\tau$ is proportional to (minus) the Hubble parameter \cite{FlavioSDbook}. The adoption of this foliation is motivated by the fact that, in it, the physical degrees of freedom of GR are spatial-conformal invariants, leading to the proposal of reformulating GR as a three-dimensional conformal field theory known as \textit{shape dynamics} \cite{FlavioSDbook,Gomes:2010fh}. The techniques employed in this work, such as the identification of the shape degrees of freedom as the physical ones, are compatible with the principles of shape dynamics, although our paper does not rely on the shape dynamical interpretation of GR. Therefore, one can view our results as a contribution to shape dynamics or as entirely independent results within Hamiltonian GR.

Now, consider the equations of motion for $x^0$ and $k_0$,
\begin{equation}
\dot x^0 = - n \, k_0 \,, \qquad
\dot k_0 = -n \left( \tfrac{1}{\sqrt{3}} \, a_0^4\, e^{\frac{2x^0}{\sqrt{3}}} \, C + \tfrac{1}{2\sqrt{3}}\, a_0^2 \, e^{\frac{x^0}{\sqrt{3}}} \, V \right) \,,
\end{equation}
and assuming, without loss of generality, $n=1$ and $a_0=1$, we can use these equations to calculate the second time derivative of the quantity $\exp\left(- x^0/ \sqrt 3 \right)$, which is a certain power of the volume:
\begin{equation}
\begin{aligned}
\frac{d^2}{dt^2} \left( e^{- \frac{x^0}{\sqrt 3}}
\right) = \frac{d}{dt} \left( e^{- \frac{x^0}{\sqrt 3}} k_0 \right) & = 
 - \tfrac{1}{3} \, e^{-\frac{x^0}{\sqrt{3}}} \left( -k_0^2 + e^{\frac{2x^0}{\sqrt{3}}} \, C \right) - \tfrac{1}{6} \, V 
\\
&\approx  \tfrac{1}{3} e^{-\frac{x^0}{\sqrt{3}}} \left( k_1^2 + k_2^2 +k_3^2 + W \right) + \tfrac{1}{6} \, V  \,, 
\end{aligned}
\end{equation}
where we used the Hamiltonian constraint in the last step. The right-hand side is non-negative because both $V$ and $W$ are positive definite. Thus, we have proven that the quantity $\exp \left( - {x^0} / {\sqrt 3}\right)$ is concave upwards. Consequently, it will  decrease monotonically for half of each solution, reaching a single minimum (which may potentially be infinitely far in time, resulting in strictly increasing or decreasing behavior), and then (if the minimum is reached in finite time) it will monotonically increase for the rest of the solution.

The volume, given by the (square root of the) inverse of this quantity, generally will monotonically increase for half of each solution, reach a maximum, and then decrease monotonically to zero. As remarked above, there may also exist degenerate solutions that undergo either monotonic growth or shrinking, reaching maximal expansion only as $t\to+ \infty$ (or $t\to -\infty$) while exhibiting a single big bang singularity as $t\to -\infty$ (or $t\to +\infty$). Our focus in this paper is solely on the behavior of the system near one singularity, while the behavior far away from it, where matter fields, cosmological constant terms, and inhomogeneities dominate the dynamics, does not concern us.

The results above represent a cosmological application of the Penrose--Hawking singularity theorems \cite{PenroseSingularityTheorem,HawkingSingularityTheorem1,HawkingSingularityTheorem2,HawkingSingularityTheorem3,HawkingEllisBook}. According to these theorems, once a solution begins to collapse, it cannot be halted and will continue to shrink until it reaches a singularity. It is important to note that, although the big bang is only reached as $t\to \pm \infty$, this does not necessarily mean that it is in the infinite future (or past). In fact, a finite amount of proper time elapses between any finite value of $t$ and $t\to \pm \infty$, as proved in \cite{Mercati_2019}. The discussion of whether this implies that the singularity is in the finite past of an observer (more precisely: in the finite past as measured by a physical clock), is subtle and was discussed in more depth in \cite{ThroughTheBigBang}.

\section[Proof of continuation under a one-dimensional \textit{ansatz}]{Proof of continuation under a one-dimensional\\\textit{ansatz}} \label{uniansatz}
\label{Sec6}

In this section, we consider a simpler but illustrative case in which the electromagnetic field has only one spatial component:
\begin{equation}
A_1 =A_2 = 0 \,, \qquad  E^1 = E^2 = 0 \,.
\end{equation}
These conditions are preserved by the equations of motion:
\begin{equation}
\begin{aligned}
\{ A_1 , \mathcal{H}[N]  \}_*\Big|_{\substack{A_1 =A_2 = 0 \\ E^1 = E^2 = 0}}   &\approx 0 \,, 
\qquad
\{ A_2, \mathcal{H}[N]   \}_*\Big|_{\substack{A_1 =A_2 = 0 \\ E^1 = E^2 = 0}}   &\approx 0 \,, 
\\
\{ E_1 ,  \mathcal{H}[N]   \}_*\Big|_{\substack{A_1 =A_2 = 0 \\ E^1 = E^2 = 0}} &  \approx 0 \,, 
\qquad
\{ E_2 ,  \mathcal{H}[N]   \}_*\Big|_{\substack{A_1 =A_2 = 0 \\ E^1 = E^2 = 0}} & \approx 0 \,, 
\end{aligned}
\end{equation}
\begin{equation}
\begin{aligned}
\{ A_1 , \mathcal{G}[\varphi]  \}_*\Big|_{\substack{A_1 =A_2 = 0 \\ E^1 = E^2 = 0}} =  0 \,, 
\qquad
\{ A_2 , \mathcal{G}[\varphi]  \}_*\Big|_{\substack{A_1 =A_2 = 0 \\ E^1 = E^2 = 0}} = 0 \,, 
\\
\{ E_1 , \mathcal{G}[\varphi]  \}_*\Big|_{\substack{A_1 =A_2 = 0 \\ E^1 = E^2 = 0}} = 0 \,, 
\qquad
\{ E_2 , \mathcal{G}[\varphi]  \}_*\Big|_{\substack{A_1 =A_2 = 0 \\ E^1 = E^2 = 0}} = 0 \,.
\end{aligned}
\end{equation}

In our \textit{ansatz}, we arbitrarily chose to keep the third component as the nonzero one, but this choice does not make the model lose any generality. In fact, if we were to choose the first or second component of $A_a$ and $E^a$ as the nonzero one, the dynamics would remain identical, with the labels for the first, second and third components permuted accordingly. This can be proven by performing a reflection transformation, such as $q_1 \to q_2$, $q_3 \to q_1$, $q_2 \to q_3$, and likewise for $p^a$. This transformation does not change the Bianchi-IX and scalar parts of the Hamiltonian constraint: this is due to a discrete symmetry of our system, which remains invariant under the aforementioned reflection transformations.

It is convenient to study how the dynamics is changed by the introduction of the one-dimensional \textit{ansatz} in terms of the original metric variables $q_a$. Therefore, we go back to \cref{GeneralBIX+GaugeHamiltonianConstraint}, and observe that  the Hamiltonian constraint reads as follows when we set $A_1 = A_2 = 0$ and $E^1 = E^2 = 0$:
\begin{equation}\label{Abelianuni}
\overline{\mathcal{H}}[N]\equiv\mathcal{H}[N]\Big|_{\substack{A_1 =A_2 = 0 \\ E^1 = E^2 = 0}} =n \left( \mathcal{H}_{BIX} +\tfrac{1}{2}(p^0)^2+  {\tfrac{1}{2}}\, q_3  \left( (E^3)^2 +  (A_3)^2 \right)  \right) \,.
\end{equation}
Now, if we consider the quantity
\begin{equation}\label{ho1d}
H^{1D}_{HO} =  (E^3)^2 +  (A_3)^2 \,,
\end{equation}
it is immediate to prove that it is conserved because it is first class with respect to the Hamiltonian constraint,
\begin{equation}
\{ H^{1D}_{HO} , \overline{\mathcal{H}}[N] \}_* \approx 0 \,,
\end{equation}
which means we can assign a constant of motion $\varepsilon$ to it, which remains unchanged throughout the solution. As a result, the geometric degrees of freedom will evolve according to the following effective Hamiltonian constraint:
\begin{equation} \label{Hconstraint}
\mathcal{H}_{ef\!f}[N]  = n \left( \mathcal{H}_{BIX} +\tfrac{1}{2}(p^0)^2+  {\tfrac{1}{2}} \, q_3  \, \varepsilon  \right) \,.
\end{equation}
We can combine the new term $ q_3\, \varepsilon/2$ with the potential term present in $\mathcal{H}_{BIX}$. For any finite value of $\varepsilon$, we can describe the dynamics of the geometrical degrees of freedom as being controlled by an effective potential given by
\begin{equation}\label{1DPotential}
U_{ef\!f} = q_1^2+q_2^2+q_3^2-\tfrac{\left(q_1+q_2+q_3\right)^2}{2}
+  {\tfrac{1}{2}}   \, q_3\,  \varepsilon  \,.
\end{equation}
Now, let us demonstrate that the additional term in the Bianchi-IX potential does not alter the result of the continuation through the singularity.

\subsection{Quiescence is unchanged}\label{subsec:quiescence}

The structure of the Hamiltonian constraint in \cref{Hconstraint} resembles that of Bianchi IX, although with a deformed potential. The solutions exhibit similar characteristics to those of Bianchi IX: stretches of inertial motion known as \textit{Kasner epochs} when the potential term is negligible and the momenta $k_1$, $k_2$, and $k_3$ (as defined at the beginning of \cref{Sec5}) are conserved. These epochs are interrupted by brief quasielastic bounces referred to as \textit{Taub transitions}. During these transitions, some of the shape momenta, namely $k_1$ and $k_2$, are dissipated \cite{FlavioSDbook}.

During a Kasner epoch, the dynamics is well approximated by that of a free particle. In these phases, the shape degrees of freedom $x^1$, $x^2$, the scale degree of freedom $x^0$, and the scalar degree of freedom $x^3$ evolve linearly with respect to parameter time $t$, so the spatial volume $v$ (defined in \cref{def of volume}) decreases exponentially. It is noteworthy that the proper time measured by a comoving observer is exponentially related to $t$, that is, it is proportional to the $t$ integral of the volume \cite{Mercati_2019,FlavioSDbook}. Therefore, if a Kasner epoch were to extend all the way to the singularity at $t \to +\infty$, only a finite amount of proper time would have elapsed \cite{Mercati_2019}. 

Conversely, in a Taub transition, the configuration point rebounds against the Bianchi-IX potential, leading to rapid changes in certain  functions of the shape variables (such as the direction of motion in configuration space). Simultaneously, $x^0$ undergoes rapid changes in speed (\textit{i.e.}, its conjugate momentum varies rapidly), but not significantly in magnitude. The resulting motion for $x^0$ is that of a segmented curve, with intervals of straight-line motion interspersed with rapid changes in slope. Proper time remains finite all the way to the singularity, namely the big bang is reached within a finite amount of proper time \cite{Mercati_2019,FlavioSDbook}.

As demonstrated in previous works, specifically \cite{ThroughTheBigBang} and \cite{Mercati_2019}, in an empty Bianchi-IX universe (\textit{i.e.}, in the absence of scalar or electromagnetic fields), the system undergoes an infinite number of Taub transitions. This chaotic behavior prevents certain degrees of freedom from converging to well-defined values at the singularity. For instance, let us consider the angular variable associated with the polar coordinates of the $(x^1,x^2)$ plane. If a Kasner epoch were to extend all the way to the singularity, this variable would eventually settle into a limiting value. However, each Taub transition makes this value change. If we were to plot its value against proper time, near the singularity it would resemble something similar to the function $\sin (1/x)$ as $x \to 0$, exhibiting an essential singularity. Consequently, it is impossible to determine the specific value that this variable takes at the big bang. This prevents any attempt to continue these solutions through it \cite{Mercati_2019,FlavioSDbook}.

\begin{figure}[!ht]\center
\includegraphics[width=0.7\textwidth]{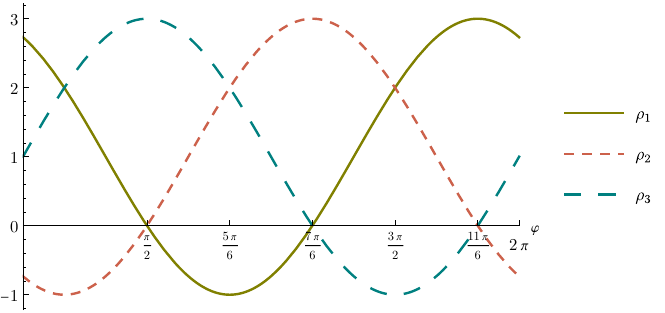} 
\captionsetup{width=.8\linewidth}
\caption{\footnotesize{Plot of the three coefficients $\rho_a$ \textit{vs.} the direction $\varphi$.}}\label{Fig:quiescence}
\end{figure}

If we introduce a scalar field, we can induce a state of \textit{quiescence} \cite{ThroughTheBigBang,Mercati_2019}, meaning that the chaotic behavior stops after a finite number of Taub bounces, and the solution settles into a last Kasner epoch lasting all the way to the big bang. However, the additional electromagnetic term in the potential \eqref{1DPotential} could, in principle, change the conditions for quiescence. We now prove that this is not the case not only for an Einstein--Maxwell-Klein--Gordon system under the one-dimensional \textit{ansatz}, but also in full generality for systems that exhibit potential terms that are polynomial in the metric components $q_1$, $q_2$, $q_3$, as is the case under our one-dimensional \textit{ansatz}. This is proven by considering the following scenario: let us temporarily remove the scalar field and assume that we begin within a Kasner epoch. Under these conditions, the Hamiltonian,  expressed in terms of the scale and shape momenta $k_0,k_1,k_2$ introduced in \cref{Sec5}, takes a simple quadratic form:
\begin{equation}\label{KasnerH}
    \mathcal{H}_\text{Kasner}=\tfrac{1}{2} \left( - k_0^2 +  k_1^2  +  k_2^2 \right)\,.
\end{equation}
The solutions are
\begin{equation}\label{KasnerSol}
x^\alpha (t) = \eta^{\alpha\beta}k_\beta \, t + x^\alpha (0) \,,
\qquad
k_\alpha (t) = v_\alpha \,,
\qquad
v_0 = +\sqrt{ (v_1)^2 + (v_2)^2 } \,,
\end{equation}
where $\eta^{\alpha\beta}=\operatorname{diag}(-1,1,1)$, $\alpha,\beta=0,1,2$. The plus sign in the dispersion relation for the integration constants $v_\alpha$ has been chosen so that the big bang singularity occurs at $t \to+ \infty$. Replacing the solution into the metric components \eqref{CanonicalTransfMisnerVariables}, we obtain

\begin{equation}\label{MetricComponentsKasnerEpoch}
\begin{aligned}
&q_1(t)=a_0^2 \, \exp \left(\tfrac{x^0 (t) -\sqrt{3}\, x^1(t)+x^2(t)}{\sqrt{3}}\right)  \propto   \exp \left(- \tfrac{\sqrt{(v_1)^2+(v_2)^2} + \sqrt{3}\, v_1 - v_2}{\sqrt{3}} \, t \right)= e^{- \rho_1 \, t} \,,
\\
&q_2(t)=a_0^2 \, \exp \left(\tfrac{x^0 (t) +\sqrt{3}\, x^1(t)+x^2(t)}{\sqrt{3}}\right)  \propto   \exp \left(- \tfrac{\sqrt{(v_1)^2+(v_2)^2} - \sqrt{3}\, v_1 - v_2}{\sqrt{3}} \, t \right)= e^{- \rho_2 \, t} \,,
\\
&q_3(t)=a_0^2 \, \exp \left(\tfrac{x^0 (t) -2\, x^2(t)}{\sqrt{3}}\right)  \propto   \exp \left(- \tfrac{\sqrt{(v_1)^2+(v_2)^2} +2\, v_2}{\sqrt{3}} \, t \right)= e^{- \rho_3 \, t}  \,.
\end{aligned}
\end{equation}
In polar coordinates, $(v_1,v_2) = |\vec v| (\cos \varphi , \sin \varphi )$, and the three coefficients $\rho_a$ appearing in the equations above can be expressed as follows:
\begin{equation}\label{PolarRhos}
\begin{aligned}
    \rho_1 &=  \tfrac{|\vec v |}{\sqrt{3}} \left(1 + \sqrt{3} \cos \varphi -  \sin \varphi \right) \,,\\
\rho_2 &=   \tfrac{|\vec v |}{\sqrt{3}} \left(1 - \sqrt{3} \cos \varphi -   \sin \varphi  \right) \,,\\
\rho_3 &=   \tfrac{|\vec v |}{\sqrt{3}} \left(1 + 2  \sin \varphi \right)\,.
\end{aligned}
\end{equation}
For any value of $\varphi$, one of the coefficients $\rho_a$ is always negative (except for the three special directions along the symmetry axes of the potential, $\varphi=\frac{\pi}{2},\frac{7\pi}{6},\frac{11 \pi}{6}$, which, however, only concern a measure-zero set of solutions). This can be observed in \cref{Fig:quiescence}. The negative $\rho_a$ coefficient will correspond to a coordinate $q_a(t) \propto \exp(- \rho_a \, t)$ which grows with $t$. As $q_a(t)$ grows, it will reach a point in which the potential term in the Hamiltonian constraint (\ref{EmptyBIXham}) will become comparable to the kinetic one, because the potential is quadratic in the variables $q_a$. When this happens, the Kasner epoch will be terminated by a Taub bounce \cite{Mercati_2019,FlavioSDbook}. Because there is always one negative $\rho_a$ parameter, a Kasner epoch cannot last forever: it will always be cut short by a bounce against the potential wall.

Now, if we reintroduce the homogeneous scalar field, the Hamiltonian constraint during a Kasner epoch \eqref{KasnerH} changes into
\begin{equation}
    \mathcal{H}_\text{Kasner}=\tfrac{1}{2} \left( - k_0^2 +  k_1^2  +  k_2^2+k_3^2 \right)\,,
\end{equation}
where we recall that $k_3$ is the conjugate momentum to the homogeneous scalar field (\ref{CanonicalTransfMisnerVariables}). A Kasner epoch in this case looks exactly the same, with the difference that the dispersion relation appearing in \eqref{KasnerSol} now looks like
\begin{equation}
v_0 = \sqrt{ (v_1)^2 + (v_2)^2 + (v_3)^2 } \,,
\end{equation}
where the constant of motion $v_3$ is the (conserved) value of $k_3$. Now, \cref{MetricComponentsKasnerEpoch} takes the same form, except that the $\rho_a$ coefficients change into 
\begin{equation}\label{Quiescent_chi}
\begin{aligned}
&\rho_1(w) =  \tfrac{|\vec v |}{\sqrt{3}} \left(w + \sqrt{3} \cos \varphi -  \sin \varphi \right) \,,
\\
&\rho_2(w) =   \tfrac{|\vec v |}{\sqrt{3}} \left(w - \sqrt{3} \cos \varphi -   \sin \varphi  \right) \,,
\\
&\rho_3(w) =   \tfrac{|\vec v |}{\sqrt{3}} \left(w + 2  \sin \varphi \right)\,,
\end{aligned} 
\qquad \quad w = \sqrt{1 + \tfrac{(v_3)^2}{ (v_1)^2 + (v_2)^2} } \,.
\end{equation}
The parameter $w$ takes the value 1 when $v_3=0$ (no scalar field), and $w>1$ when $v_3 \neq 0$. Each Taub transition ends in a new Kasner epoch with a lower value of $(v_1)^2 + (v_2)^2$ (see \cite{Mercati_2019,FlavioSDbook} for the proof), so the parameter $w$ progressively grows larger after each bounce. When it reaches values equal to or larger than $w=2$, all the $\rho_a(w)$ functions become positive everywhere. We reach a situation in which all the terms in any potential that is polynomial in $q_a$ can only decrease with time. Of course $w$ does not need to reach the value $w=2$ for the system to settle around one last Kasner epoch: any value $w >1$ allows for intervals of values of the $\varphi$ angle in which all the $\rho_a$ functions are positive, and, if the solution evolves sufficiently parallel to one of those windows, it will never exit it and thereby achieve quiescence. When this happens, the solution settles with increasing accuracy around a single Kasner epoch all the way to the singularity, without further Taub bounces.

As we mentioned earlier, the effective potential of the one-dimensional model \eqref{1DPotential} is polynomial in $q_a$ (it includes quadratic terms from the Bianchi-IX part (\ref{EmptyBIXham}) and a linear term in $q_3$). Therefore, the conditions for quiescence remain completely unchanged with respect to the model without gauge fields considered in \cite{ThroughTheBigBang}.

However, it is important to notice that the polynomiality of the potential is not guaranteed in general. From \cref{GeneralBIX+GaugeHamiltonianConstraint}, we can observe that in the general case where the electromagnetic field has more than one spatial component, there are nonpolynomial terms, such as $q_1q_2 / (q_1-q_2)$, and so on.

\subsection{Continuing the dynamics through the singularity} \label{subsec:continuation1d}

In the previous section, we proved that the presence of a one-dimensional electromagnetic field does not alter the quiescent behavior as the system approaches the big bang. This provides the foundation for extending the continuation result of \cite{ThroughTheBigBang,Mercati_2019} to GR minimally coupled with electromagnetism under the one-dimensional \textit{ansatz}.

It is important to note that the variables $x^0$, $x^1$, $x^2$, $x^3$, $k_0$, $k_1$, $k_2$, $k_3$ are not a suitable set for describing the system at the big bang. For example, the singularity is located at the boundary of the $(x^1,x^2)$ plane, where $(x^1)^2 + (x^2)^2 \to \infty$. Therefore, when expressed in terms of $x^0,k_0, \dots, x^3,k_3$, the solutions become degenerate at the big bang. The values of certain variables (such as $(x^1)^2 + (x^2)^2$) at the singularity do not depend on the choice of initial values and cannot be, in this sense, predictive. However, we can demonstrate that this loss of predictability at the big bang is coordinate dependent. It is possible to find a sufficiently large set of variables that tend to finite \textit{nontrivial} limits at the big bang, and at the same time possess the property that specifying their values at any instant, including at the singularity, uniquely determines the solution.
 
Specifically, we can demonstrate that the equations of motion in these variables form an autonomous set of ordinary differential equations (ODEs) that are regular at the big bang. This means that the right-hand sides of the equations of motion tend to finite limits, as do their first derivatives. At the singularity, these equations satisfy the conditions required by the Picard-Lindel\"of theorem of existence and uniqueness of solutions of ODEs \cite{teschl2012ordinary}. Thus, it is possible to set an initial value problem at the big bang that has a unique solution. Consequently, the big bang is not a region where determinism fails, as \textit{no information about the dynamical system is lost there.}

\begin{figure}[!ht]
    \centering
    \begin{subfigure}[b]{\textwidth}
        \centering
        \includegraphics[height=0.28\textheight]{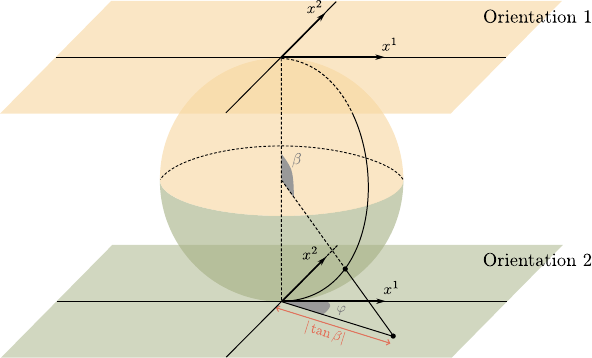}
        \subcaption{}
        \label{FigGnomonic1}
    \end{subfigure}
    \\
    \centering
    \vspace{10pt}
    \begin{subfigure}[b]{\textwidth}
        \centering
        \includegraphics[height=0.28\textheight]{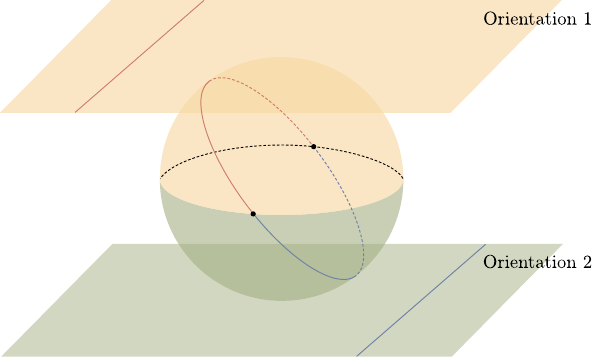}
        \subcaption{}
        \label{FigGnomonic2}
    \end{subfigure}
    \captionsetup{width=.8\linewidth}
    \caption{\footnotesize{Gnomonic map in two dimensions. (a) Gnomonic projection from (two copies of) a plane to a sphere. (b) Straight lines on planes are projected onto half great circles on the gnomonic sphere.}}
\end{figure}

If the singularity is a region where the existence and uniqueness theorem holds, a unique solution should depart from any of its points, \textit{in two directions}. One direction leads to the interior of the configuration space we used so far. However, it is not clear at this point where the other direction should lead. In fact, the singularity lies at the boundary of the configuration space, and we need to extend this space in order to discuss the fate of the solutions that reach the big bang. The aforementioned regular variables enable us to achieve such an extension in a natural manner: the shape and scalar variables $x^1$, $x^2,$ $x^3$ are related to the three regular variables $\beta$, $\theta$, $\varphi$ through a \textit{gnomonic map} \cite{ThroughTheBigBang,FlavioSDbook},
\begin{equation}
\left\{
\begin{aligned}
& x^1 = |\tan \beta| \, \sin\theta \, \cos \varphi \,, \\
& x^2 = |\tan \beta| \, \sin\theta \, \sin \varphi \,, \\
& x^3 = |\tan \beta| \, \cos\theta \,,
\end{aligned}
\right. 
\end{equation}
where $\beta,\theta \in [0, \pi]$, $\varphi\in [0,2\pi)$ are hyperspherical coordinates on a three-sphere. These coordinates project the configuration space $(x^1,x^2,x^3)$ onto a hemisphere of a three-sphere. The gnomonic map defines a double cover of a three-dimensional plane by a three-sphere (see \cref{FigGnomonic1}), in which each hemisphere is mapped to a $(x^1,x^2,x^3)$ hyperplane, extending the original shape space into two copies of itself. Physically, we interpret each hyperplane as the shape space of a three-geometry plus a scalar field (recall that the shape space includes only the anisotropy and matter degrees of freedom, and excludes the volume/scale one; see \cref{Sec5}), with a fixed spatial orientation. Each hyperplane has a different orientation, that flips upon crossing the boundary between the two hyperplanes (the equator of the three-sphere) \cite{ThroughTheBigBang,Mercati_2019}. In \cite{ThroughTheBigBang}, solutions that cross the boundary between the two fixed-orientation halves of shape space were interpreted as a universe that approaches the singularity with a certain spatial orientation,  and collapses, at the big bang, into a degenerate zero-volume one-dimensional geometry, in which two spatial directions are infinitely smaller than the third one.\footnote{There is also a measure-zero set of solutions of two-dimensional degenerate geometries, in which one direction is infinitely smaller than the other two \cite{Mercati_2019}.} Once the big bang is crossed, the volume can start growing again, but a universe with an opposite spatial orientation will emerge \cite{ThroughTheBigBang,Mercati_2019}. This entire process can be described using the extended shape space (namely, the gnomonic three-sphere) where the singularity is projected from the boundary of the plane associated with a fixed spatial orientation onto the equator of the sphere ($\beta = \tfrac{\pi}{2}$). Therefore, the big bang is approached as $\beta\to \tfrac{\pi}{2}^\pm$, while the angles $\theta$ and $\varphi$ represent the direction in which the equator is approached in the extended configuration space. Quiescent solutions, which were straight lines in the configuration plane, are projected onto half great circles on the gnomonic sphere. Each half great circle has a unique natural and regular continuation, which corresponds to the other half of the same great circle in the other hemisphere (see \cref{FigGnomonic2}).

The gnomonic projection suggests a natural continuation rule for purely Kasner solutions: they are great circles on the gnomonic sphere \cite{ThroughTheBigBang}, which correspond to two (generally distinct) straight lines on the two $(x^1, x^2,x^3)$ hyperplanes associated with the two spatial orientations (see \cref{FigGnomonic2}). To extend this result to Bianchi-IX solutions, where the straight lines only exist in a neighborhood of the singularity, we must identify an additional set of five variables that exhibit a finite nontrivial value at the singularity.

The shape and scalar conjugate momenta $k_1$, $k_2$, $k_3$ can be regularized through the following change of variables:
\begin{equation}
\left\{
\begin{aligned}
J &= \operatorname{sgn}(\tan \beta)\, \tfrac{x^1k_1+x^2k_2+x^3k_3}{\sqrt{(x^1)^2+(x^2)^2+(x^3)^2}}  \,, \\
L_1 &= x^2k_3-x^3k_2  \,, \\
L_3 &= x^1k_2-x^2k_1   \,,
\end{aligned}
\right. 
\end{equation}
while the scale variable and its conjugate momentum, $x^0$ and $k_0$, require the following transformation:
\begin{equation}
\left\{
\begin{aligned}
& \eta = \operatorname{sgn} (\tan\beta) \left(x^0 + k_0 \, \tfrac{(x^1)^2+(x^2)^2+(x^3)^2}{x^1k_1+x^2k_2+x^3k_3} \right) \,, \\
& \kappa = |k_0| \,.
\end{aligned}
\right.
\end{equation}
The variables we introduced tend to a finite limit as the singularity is approached by a quiescent solution:
\begin{equation}
\left\{
\begin{aligned}
J &\to   \sin\theta\, (v_1 \cos\varphi + v_2 \sin\varphi) + v_3 \cos\theta \,, \\
L_1 &\to  \tan\beta\, (v_3 \sin\theta\sin\varphi - v_2 \cos\theta) \,, \\
L_3 &\to  \tan\beta\, \sin\theta\, (v_2 \cos\varphi - v_1 \sin\varphi) \,, \\
\eta &\to   \operatorname{sgn} (\tan\beta) \, x^0 + v_0\, J^{-1} \tan\beta \,, \\
\kappa &\to  v_0 \,,
\end{aligned}
\right. \qquad \text{as} \qquad \beta \to \tfrac{\pi}{2}^\pm \,.
\end{equation}
The solution identified by the initial data $v_0$, $v_1$, $v_2$, $v_3$, $x^0(0)$, $x^1(0)$, $x^2(0)$, $x^3(0)$ can be matched to a unique solution belonging to the other hemisphere with initial data $-v_0$, $-v_1$, $-v_2$, $-v_3$, $-x^0(0)$, $-x^1(0)$, $-x^2(0)$, $-x^3(0)$. It is worth noting that the second solution reaches the limit as $t \to -\infty$, \textit{i.e.}, the big bang singularity of the universe with the opposite spatial orientation is reached as $t \to -\infty$.

We have yet to discuss the electromagnetic degrees of freedom. Under the one-dimensional \textit{ansatz}, there are two electromagnetic variables, $A_3$ and $E^3$. Their equations of motion with respect to the effective Hamiltonian \eqref{Hconstraint} (for $n=1$) are the following:
\begin{equation}
\dot A_3 = q_3 \, E^3 \,,
\qquad 
\dot E^3= - q_3 \, A_3 \,.
\end{equation}
At the singularity all the metric variables $q_a$ go to zero (see \cref{Sec6}), so $A_3$ and $E^3$ become conserved as we approach the big bang. The electromagnetic variables are unaffected by the orientation flip, meaning that the constant values to which these variables tend are the same regardless of whether the singularity is approached from the left or the right ($\beta\to\tfrac{\pi}{2}^{+}$ or $\beta\to\tfrac{\pi}{2}^{-}$). Thus, these variables are already regular and effectively describe the evolution of electromagnetic degrees of freedom in the entire extended configuration space.

Expressed in terms of the variables $\beta$, $\theta$, $\varphi$, $\eta$, $J$, $L_1$, $L_3$, $\kappa$, $A_3$, $E^3$, and assuming the quiescence conditions are satisfied (\textit{i.e.}, neglecting the potential terms), the Hamiltonian constraint \eqref{Hconstraint} becomes
\begin{equation}\label{HK}
    \mathcal{H}_{Kasner} = \tfrac{1}{2} \left( \kappa^2 - J^2 - \tfrac{L_1^2}{\tan^2\beta \sin^2\varphi} + \tfrac{L_3^2\left( \cos^{2}\varphi - \sin^{-2}\theta \right)}{\tan^2\beta \sin^2\varphi} - \tfrac{L_1 L_3}{\tan^2\beta \tan \theta \tan \varphi \sin\varphi} \right) \,,
\end{equation}
where we set $n=1$ for sake of simplicity.

The equations of motion for the variables $\beta$, $\theta$, $\varphi$, $\eta$, $J$, $L_1$, $L_3$, $\kappa$, $A_3$, $E^3$ with respect to the coordinate time $t$ can be obtained by calculating the Dirac brackets using the Hamiltonian constraint \eqref{Hconstraint}. However, $t$ is not a suitable choice of independent variable at the singularity as it diverges there. Instead, a natural choice of independent variable is the arc length on the gnomonic sphere:
\begin{equation}
d\ell = \sqrt{d\beta^2 + \sin^2\beta (d\theta^2+\sin^2\theta \, d\varphi^2)} \,,
\end{equation}
which is automatically monotonic everywhere on a solution and tends to a  finite limit at the singularity.
The equations of motion give us the relationship between the two independent variables:
\begin{equation}
\tfrac{d \ell}{d t} =  \Lambda^{-1} \cos^2\beta \,, \qquad
\Lambda  = \left(J^2 + \sin^{-2} \beta \left( \left(L_3 \sin\theta\right)^2 + \left( \tfrac{L_1}{\sin\varphi} + \tfrac{L_3}{\tan\theta\tan\varphi} \right)^2 \right)\right)^{-\tfrac{1}{2}} \, .
\end{equation}

The equations of motion with respect to the arc length during quiescence read
\begin{equation} \label{eom}
\begin{array}{l l}
\begin{aligned}
\tfrac{d\beta}{d\ell} &= \Lambda \,J \,, \\
\tfrac{d\theta}{d\ell} &= - \Lambda \left( L_1 + L_3 \tfrac{\cos \varphi}{\tan \theta }\right)\sin^{-2} \beta \sin^{-1} \varphi \,, \\
\tfrac{d\varphi}{d\ell} &= \Lambda \, L_3 \sin^{-2} \beta \sin^{-2} \theta \,,  \\
\tfrac{d\eta}{d\ell} &= - \Lambda \, \Theta \,\kappa \, J^{-2} \sin^{-2} \beta \,, \\
\tfrac{dA_3}{d\ell} &= 0 \,,
\end{aligned} &
\begin{aligned}
\tfrac{dJ}{d\ell} &= \Lambda \, \Theta \, \cos \beta \sin^{-3} \beta \,, \\
\tfrac{dL_1}{d\ell} &= 0 \,, \\ 
\tfrac{dL_3}{d\ell} &= 0 \,,  \\
\tfrac{d\kappa}{d\ell} &= 0 \,, \\
\tfrac{dE^3}{d\ell} &= 0 \,,
\end{aligned}
\end{array}
\end{equation}
where
\begin{equation}
\Theta  = \left( \tfrac{L_3^2}{\tan^{2} \theta \tan^{2} \varphi} + \tfrac{L_3^2}{\sin^{3} \theta} + \tfrac{2 L_1 L_3}{\tan \theta \tan \varphi \sin \varphi} + \tfrac{L_1^2}{\sin^{2}\varphi} \right) \sin \theta \,.
\end{equation}
It is important to notice that this model has ten degrees of freedom, but only eight of them are truly physical, as two are redundant due to the Hamiltonian constraint and its gauge fixing. To eliminate the two remaining non physical degrees of freedom, we need to solve the Hamiltonian constraint \eqref{HK} with respect to one of the variables and impose a gauge-fixing condition. A straightforward choice for gauge fixing the Hamiltonian constraint, which also serves as the generator of the dynamics, is to fix a specific instant of time. In our case, the natural choice is $\beta = \tfrac{\pi}{2}$, which represents the instant of the singularity. The suitability of fixing $\beta$ as a gauge for \eqref{HK} can be verified by calculating the Dirac bracket between $\beta$ and \eqref{HK} and observing that it is never zero at $\beta = \tfrac{\pi}{2}$.

At the singularity, the Hamiltonian constraint tends to the simple expression
\begin{equation}
    \mathcal{H}_{Kasner}=\frac{1}{2} (\kappa^2 - J^2) \,.
\end{equation}
Therefore, it can be easily solved with respect to either the variable $\kappa$ or $J$. Once we compute the equations of  motion and their first derivatives, we can impose the condition $\kappa = J$ (keeping in mind that both $\kappa$ and $J$ are positive definite at the singularity) if we wish to eliminate this last redundant degree of freedom.

We are now prepared to present the continuation result. The equations of motion \eqref{eom} are regular at the singularity, meaning that they admit the same left and right limit as $\beta \to \tfrac{\pi}{2}$:
\begin{equation}
\begin{array}{l l}
\begin{aligned}
\tfrac{d\beta}{d\ell} & \to \Lambda_{\pi / 2} \,J \,, \\
\tfrac{d\theta}{d\ell} & \to - \Lambda_{\pi / 2} \left( L_1 + L_3 \tfrac{\cos \varphi}{\tan \theta }\right) \sin^{-1} \varphi \,, \\
\tfrac{d\varphi}{d\ell} & \to \Lambda_{\pi /2} \, L_3 \sin^{-2} \theta \,,  \\
\tfrac{d\eta}{d\ell} & \to - \Lambda_{\pi / 2} \, \Theta \,\kappa \, J^{-2} \,, \\
\tfrac{dA_3}{d\ell} &\to 0 \,,
\end{aligned} &
\begin{aligned}
\tfrac{dJ}{d\ell} &\to 0 \,, \\
\tfrac{dL_1}{d\ell} & \to 0 \,, \\ 
\tfrac{dL_3}{d\ell} & \to 0 \,,  \\
\tfrac{d\kappa}{d\ell} & \to 0 \,, \\
\tfrac{dE^3}{d\ell} & \to 0 \,,
\end{aligned}
\end{array}
\end{equation}
where:
\begin{equation}
\Lambda_{\pi / 2} = \lim_{\beta\to \pi/2} \Lambda = \left(J^2 +  \left(L_3 \sin\theta\right)^2 + \left( \tfrac{L_1}{\sin\varphi} + \tfrac{L_3}{\tan\theta\tan\varphi} \right)^2 \right)^{-\tfrac{1}{2}} \,.
\end{equation}
The assumptions of the Picard--Lindel\"of theorem require the Lipschitz continuity of the right-hand side of the equations of motion. However, in our case, we can prove an even stronger condition: differentiability.
In fact, the first derivatives of the right-hand sides of \eqref{eom} with respect to all the variables $\beta$, $\theta$, $\varphi$, $\eta$, $J$, $L_1$, $L_3$, $\kappa$, $A_3$, $E^3$ are all regular as $\beta \to \tfrac{\pi}{2} ^{\pm}$:

\begin{equation}
\begin{aligned}\label{FirstDerivatives}
\partial_\beta \left(\tfrac{dJ}{d\ell}\right) & \to - \Lambda_{\pi /2} \, \Theta \,, \\
\partial_\theta \left(\tfrac{d\beta}{d\ell}\right) & \to - \Lambda_{\pi /2}^3  \, J \sin \theta \left( L_3^2 \cos \theta - \tfrac{L_3^2}{\tan \theta \tan^2\varphi \sin^3 \theta} - \tfrac{L_1 L_3}{\sin \varphi \tan\varphi \sin^3 \theta} \right) \,, \\
\partial_\theta \left(\tfrac{d\theta}{d\ell}\right) & \to \Lambda_{\pi /2}^3 \, \tfrac{L_3}{\sin^2\theta} \left( \tfrac{L_1 L_3 \cos \theta \sin^3 \theta}{\sin \varphi} + \tfrac{J^2 + \tfrac{1}{2} L_3^2 \sin^2\theta \, (3+ \cos (2\theta))}{\tan \varphi}\right) \,, \\
\partial_\theta \left(\tfrac{d\varphi}{d\ell}\right) & \to \Lambda_{\pi /2}^3 \,  \tfrac{L_3}{\sin^2\theta} \biggl( \tfrac{L_3^2}{\tan\theta \tan^2\varphi \sin^2\theta } + \tfrac{L_1 L_3}{ \tan\varphi \sin^2\theta \sin \varphi }  - \tfrac{2 J^2}{\tan\theta}\biggr.\\
& \qquad \qquad \qquad ~~\biggl. - \tfrac{2}{\tan\theta} \left( \tfrac{L_1}{\sin\varphi} + \tfrac{L_3}{\tan\theta \tan \varphi} \right)^2 - 3\,L_3^2 \cos\theta \sin\theta \biggr) \,, \\ 
\partial_\theta \left(\tfrac{d\eta}{d\ell}\right) & \to \Lambda_{\pi /2}^3\, \tfrac{\kappa}{J^2}  \biggl( -L_3 \, \Theta \, \sin\theta \left( \tfrac{L_3}{\tan\theta \tan^2\varphi \sin^3\theta} + \tfrac{L_1}{\tan\varphi \sin^3\theta \sin\varphi}-L_3\cos\theta \right) \biggr. \\
& \qquad \qquad \qquad ~~ \biggl. + \Lambda_{\pi /2}^{-2} \, \tfrac{L_3}{\sin\theta} \left( \tfrac{L_3}{\tan\theta} \left( \tfrac{2}{\tan^2\varphi} + \tfrac{3}{\sin\theta} \right) + \tfrac{2L_1}{\tan\varphi \sin\varphi} \right) - \Lambda_{\pi /2}^{-2} \, \Theta \cos \theta \biggr) \,, \\
\partial_\varphi \left(\tfrac{d\beta}{d\ell}\right) & \to \Lambda_{\pi /2}^3 \, \tfrac{J}{\sin\varphi} \left( \tfrac{L_3}{\tan\theta\tan\varphi} + \tfrac{L_1}{\sin\varphi}\right) \left( \tfrac{L_1}{\tan\varphi} +\tfrac{L_3}{\tan\theta\sin\varphi} \right) \,, \\
\partial_\varphi \left(\tfrac{d\theta}{d\ell}\right) & \to \Lambda_{\pi /2}^3 \, \tfrac{2 J^2  + L_3^2 + L_3^2 \cos(2\theta)}{2\sin^2\varphi} \left( L_1 \cos\varphi + \tfrac{L_3}{\tan\theta} \right) \,, \\
\partial_\varphi \left(\tfrac{d\varphi}{d\ell}\right) & \to \Lambda_{\pi /2}^3 \, \tfrac{L_3}{\sin^2\theta \sin\varphi} \left( \tfrac{L_3}{\tan\theta\tan\varphi} + \tfrac{L_1}{\sin\varphi} \right) \left( \tfrac{L_1}{\tan\varphi}+ \tfrac{L_3}{\tan\theta \sin\varphi} \right) \,, \\
\partial_\varphi \left(\tfrac{d\eta}{d\ell}\right) & \to \Lambda_{\pi /2}^3 \, \tfrac{\kappa \sin^3\theta}{J^2 \sin^2\varphi} \left( \tfrac{2L_1^2}{\tan\varphi} + \tfrac{2L_3^2}{\tan\varphi \tan^2\theta} + \tfrac{L_1L_3 \left( 3+\cos (2\theta) \right)}{\tan\theta\sin\varphi} \right) \\
& \qquad \qquad \qquad\qquad ~~ \left( 2L_3^2-\tfrac{L_3^2}{\sin^5\theta} + \tfrac{2J^2}{\sin^2\theta} + \left( \tfrac{L_3}{\tan\theta\tan\varphi\sin\theta} + \tfrac{L_1}{\sin\varphi\sin\theta} \right)^2 \right) \,, \\
\partial_J \left(\tfrac{d\beta}{d\ell}\right) & \to \Lambda_{\pi /2}^3 \, \left( \left(L_3 \sin\theta\right)^2 + \left( \tfrac{L_1}{\sin\varphi} + \tfrac{L_3}{\tan\theta\tan\varphi} \right)^2 \right) \,, \\
\partial_J \left(\tfrac{d\theta}{d\ell}\right) & \to \Lambda_{\pi /2}^3 \, \tfrac{J^2}{\sin\varphi} \left( L_1+\tfrac{L_3\cos\varphi}{\tan\theta} \right) \,, \\
\partial_J \left(\tfrac{d\varphi}{d\ell}\right) & \to -\Lambda_{\pi /2}^3 \, \tfrac{J L_3}{\sin^2\theta} \,, \\
\partial_J \left(\tfrac{d\eta}{d\ell}\right) & \to \Lambda_{\pi /2}^3 \, \Theta \, \tfrac{\kappa}{J^3} \left( 3J^2 +2\left(L_3 \sin\theta\right)^2 + 2\left( \tfrac{L_1}{\sin\varphi} + \tfrac{L_3}{\tan\theta\tan\varphi} \right)^2 \right)\,, \\
\partial_{L_1} \left(\tfrac{d\beta}{d\ell}\right) & \to -\Lambda_{\pi /2}^3 \, \tfrac{J}{\sin\varphi} \left( \tfrac{L_3}{\tan\theta\tan\varphi} + \tfrac{L_1}{\sin\varphi}\right) \,, \\
\partial_{L_1} \left(\tfrac{d\theta}{d\ell}\right) & \to -\Lambda_{\pi /2}^3 \, \tfrac{J^2 + L_3^2\sin^2\theta}{\sin\varphi} \,, \\
\partial_{L_1} \left(\tfrac{d\varphi}{d\ell}\right) & \to -\Lambda_{\pi /2}^3 \, \tfrac{L_3}{\sin^2\theta \sin\varphi} \left( \tfrac{L_3}{\tan\theta\tan\varphi} + \tfrac{L_1}{\sin\varphi}\right) \,, \\
\partial_{L_1} \left(\tfrac{d\eta}{d\ell}\right) & \to - \Lambda_{\pi /2}^3 \, \tfrac{\kappa \sin^3\theta}{J^2 \sin\varphi} \left( \tfrac{L_3}{\tan\theta\tan\varphi} + \tfrac{L_1}{\sin\varphi}\right) \\
& \qquad \qquad \qquad\qquad ~~ \left( 2L_3^2 - \tfrac{L_3^2}{\sin^5\theta} + \tfrac{2J^2}{\sin^2\theta} + \left( \tfrac{L_3}{\tan\theta\tan\varphi\sin\theta} + \tfrac{L_1}{\sin\varphi\sin\theta} \right)^2  \right) \,,\\
\partial_{L_3} \left(\tfrac{d\beta}{d\ell}\right) & \to - \Lambda_{\pi /2}^3 \, J \left( L_3\sin^2\theta + \tfrac{L_3}{\tan^2\theta\tan^2\varphi} + \tfrac{L_1}{\tan\theta\tan\varphi \sin\varphi} \right) \,, \\
\partial_{L_3} \left(\tfrac{d\theta}{d\ell}\right) & \to \Lambda_{\pi /2}^3 \, \left( \tfrac{L_1L_2\sin^2\theta}{\sin\varphi} - \tfrac{J^2}{\tan\theta\tan\varphi} \right) \,, \\
&\hspace{5.12in}\llap{\text{(continued on next page)}}
\end{aligned}
\end{equation}
\begin{equation*}
\begin{aligned}
&\hspace{5.12in}\llap{\text{(continued from previous page)}}\\
\partial_{L_3} \left(\tfrac{d\varphi}{d\ell}\right) & \to \Lambda_{\pi /2}^3 \, \left( \tfrac{J^2}{\sin^2\theta} + \tfrac{L_1L_3}{\tan\theta\tan\varphi\sin\varphi\sin^2\theta} + \tfrac{L_1^2}{\sin^2\theta\sin^2\varphi} \right) \,, \\
\partial_{L_3} \left(\tfrac{d\eta}{d\ell}\right) & \to  \Lambda_{\pi /2}^3 \, \tfrac{\kappa}{J^2} \biggl( \left( \tfrac{L_3\cos\theta}{\tan\theta\tan^2\varphi} + \tfrac{L_1\cos\theta}{\tan\varphi\sin\varphi} \right) \left( 2J^2 + \left( \tfrac{L_3}{\tan\theta\tan\varphi} + \tfrac{L_1}{\sin\varphi} \right)^2 \right) \biggr.\\
& \qquad \qquad \qquad ~~ -  \tfrac{L_3^3\cos^2\theta\sin\theta}{\tan^2\varphi} + \tfrac{L_1^2L_3\sin^3\theta}{\sin^2\varphi} - L_3^3 \\
& \qquad \qquad \qquad ~~ - \biggl.  \tfrac{L_3}{\sin^2\theta} \left( 2J^2 + \left( \tfrac{L_3}{\tan\theta\tan\varphi} + \tfrac{L_1}{\sin\varphi} \right) \left( \tfrac{L_3}{\tan\theta\tan\varphi} + \tfrac{2L_1}{\sin\varphi} \right) \right)  \biggr) \,, \\
\partial_\kappa \left(\tfrac{d\theta}{d\ell}\right) & \to - \Lambda_{\pi /2}\, \Theta \, J^{-2} \,, 
\end{aligned}
\end{equation*}
and all the other derivatives tend to zero. Notice  that imposing the asymptotic solution of the Hamiltonian constraint, $\kappa = J$, does not alter the regularity of the right-hand sides.

In full generality, when the potential terms cannot be neglected, the Hamiltonian constraint \eqref{Hconstraint} assumes the following form in the new variables:
\begin{equation}
\begin{aligned}\label{Hfull}
\mathcal{H}_{eff} & = \mathcal{H}_{Kasner} + \tfrac{1}{2} e^{\tfrac{2}{\sqrt{3}} \operatorname{sgn} (\tan\beta)\left( \eta-\kappa\, J^{-1} \tan\beta \right)} C(\beta,\theta,\varphi) \\
& \qquad \qquad \qquad ~~ + \tfrac{1}{2} e^{\tfrac{1}{\sqrt{3}} \operatorname{sgn} (\tan\beta)\left( \eta-\kappa\, J^{-1} \tan\beta \right)} e^{-\tfrac{2}{\sqrt{3}} |\tan\beta| \sin\theta\sin\varphi} \varepsilon \,,
\end{aligned}
\end{equation}
where $C(\beta,\theta,\varphi)$ represents the Bianchi-IX potential \eqref{BIXpot} as a function of $\beta,\theta,\varphi$. When the quiescent approximation is relaxed, the equations of motion \eqref{eom} acquire additional ``force'' terms arising from the potential. However, these terms are strongly suppressed near the equator/singularity, due to the exponential factors in \cref{Hfull}, which tend to zero as $\beta \to \tfrac{\pi}{2}$ like $\exp \left(- \text{\it const.} |\tan\beta| \right)$ (after solving the Hamiltonian, \textit{e.g.}, with respect to $\kappa$, and substituting the solution back into the equations of motion). In the equations of motion, the suppressing exponentials appear multiplied by powers of $\tan \beta$. Although the positive powers diverge, they do so slower than the exponentials and end up suppressed as well. As a result, the full equations of motion asymptotically tend to the quiescent ones \eqref{eom}. This holds true for the first-derivative expressions \eqref{FirstDerivatives} as well, once again due to the presence of the suppressing exponentials.

\section{Extension of the proof to generic electromagnetic fields}
\label{Sec7}

In the most generic situation, the electromagnetic field has all three components. The Hamiltonian constraint is given by \cref{H3d}. In this case as well, we can identify a conserved quantity (\textit{i.e.}, one that is first class with respect to the Hamiltonian constraint):
\begin{equation}\label{3DconsQuant}
     H^{3D}_{HO} = \sum_{a=1}^3 \left( (E^a)^2 + (A_a)^2 \right) \,, \qquad \{ H^{3D}_{HO} , \mathcal{H}[N] \}_* = 0 \,. 
\end{equation}
This can be readily proven by observing that $\mathcal{H}[N]$ depends on the electromagnetic variables only though the six terms $(E^a)^2+(A_a)^2$ and $M_a$ (the latter defined in \cref{DefinizioneVettoreM}), and each of these terms commutes separately with $H^{3D}_{HO}$. These six terms correspond to the conserved quantities of a three-dimensional Harmonic oscillator, namely, three ``energies'' and three components of the angular momentum. Hence, we can associate a constant of motion $\varepsilon_{3D}$ to $H^{3D}_{HO}$. It is important to notice that, although this quantity is not explicitly present in \eqref{H3d}, it establishes bounds on the possible values of the electromagnetic field and momenta:
\begin{equation}
| A_a |   \leq  \sqrt{\varepsilon_{3D}} \,, \qquad  | E^a |   \leq  \sqrt{\varepsilon_{3D}} \,.
\end{equation}

We now demonstrate that the conditions for quiescence are still satisfied, even without the one-dimensional \textit{ansatz} for the electromagnetic field. As mentioned before, the relevant Hamiltonian constraint in this scenario is given by \cref{H3d}. Since the electromagnetic variables only appear in the potential term $U(x,A,E)$, the removal of the one-dimensional \textit{ansatz} does not affect the results obtained in \cref{subsec:quiescence}: during a Kasner epoch, the metric components $q_a$ progressively decrease in time, along with any polynomial quantity derived from them. However, in the absence of the one-dimensional \textit{ansatz}, the potential $U$ acquires two additional terms (see Eqs.~\eqref{BIXpot} and \eqref{VandWpots}), one of which is not even polynomial in $q_a$. Consequently, they need separate discussion.

The potential $U$ now consists of a combination of three quantities: $C$, $V$, and $W$. The Bianchi-IX potential $C(x^1,x^2)$ depends only on the metric variables and is polynomial in $q_a$, hence its behavior is analogous to that of the effective potential in the one-dimensional model \eqref{1DPotential}. The potential $V(x^1,x^2,A,E)$ is again polynomial in $q_a$, but its coefficients are functions of the electromagnetic variables. The presence of the conserved quantity \eqref{3DconsQuant} implies that $E^a$ and $A_a$ can only oscillate within finite and fixed values, thus the behavior of $V$ is controlled by that of $q_a$. Similarly, the behavior of $W(x^1,x^2,A,E)$ is determined by $q_a$. However, in this case, the dependence of $W$ on $q_a$ is nonpolynomial. $W$ depends on the following three functions of $q_a$:
\begin{equation}\label{NonPolyObjects}
    \begin{gathered}
        \frac{q_1 q_2}{(q_1-q_2)^2} = \frac{e^{2 |\vec{v}| t \cos \varphi}}{(e^{2 |\vec{v}| t \cos \varphi}-1)^2} \, , \qquad
        \frac{q_2 q_3}{(q_2-q_3)^2} = \frac{e^{|\vec{v}| t (\cos \varphi+\sqrt{3}\sin\varphi)}}{(e^{|\vec{v}| t (\cos \varphi+\sqrt{3}\sin\varphi)}-1)^2} \, , \\
        \frac{q_3 q_1}{(q_3-q_1)^2} = \frac{e^{|\vec{v}| t (\cos \varphi+\sqrt{3}\sin\varphi)}}{(e^{|\vec{v}| t \cos \varphi}-e^{\sqrt{3} |\vec{v}| t \sin \varphi})^2} \, , \\
    \end{gathered}
\end{equation}
where $q_a$ has been replaced by the solutions of the equations of motion during a Kasner epoch, as given by \cref{MetricComponentsKasnerEpoch}, with the velocities expressed in polar coordinates as in \cref{PolarRhos}. As $t \to + \infty$, the three quantities in \cref{NonPolyObjects} tend to zero for all values of $\varphi$, except $\varphi = \frac{\pi}{6}, \frac{\pi}{2}, \frac{5\pi}{6}, \frac{7\pi}{6}, \frac{3\pi}{2}, \frac{11\pi}{6}$. These six directions are parallel to the three axes of symmetry of the shape potential $C(x_1,x_2)$ \cite{ThroughTheBigBang,Mercati_2019}. Along these directions, two of the metric components $q_a$ are identical, and one of the quantities in \cref{NonPolyObjects} becomes infinite. This singularity only affects a measure-zero set of solutions (those confined along the symmetry axes), which require a different gauge fixing of the diffeomorphism constraint, and therefore need to be described with a different set of variables. Their continuability can be discussed separately, and we are not interested in special sets of solutions in the present paper, so we ignore them for now.

We have demonstrated that the removal of the one-dimensional \textit{ansatz} does not hinder quiescence: all the potential terms decrease with time, allowing the solution to settle around a single Kasner epoch all the way to the singularity.

Having established that the entire system (\textit{i.e.}, including all six electromagnetic degrees of freedom) exhibits quiescent behavior as it approaches the big bang, we now proceed to demonstrate that the continuation result holds as well. To prove this, we follow the same procedure as described in \cref{subsec:continuation1d}. In terms of the variables $\beta$, $\theta$, $\varphi$, $\eta$, $J$, $L_1$, $L_3$, $\kappa$, $A_1$, $E^1$, $A_2$, $E^2$, $A_3$, $E^3$, the dynamics governed by the Hamiltonian constraint \eqref{H3d} is indistinguishable from that generated by \eqref{Hconstraint} when the quiescence conditions are satisfied. Therefore, the equations of motion can be well approximated by \cref{eom}, with the addition of
\begin{equation}
      \tfrac{dA_1}{d\ell}  = 0 \,, \qquad
      \tfrac{dA_2}{d\ell}  = 0 \,, \qquad
      \tfrac{dE^1}{d\ell}  = 0 \,, \qquad
      \tfrac{dE^2}{d\ell}  = 0 \,,
\end{equation}
whose right-hand sides are differentiable, similar to the other equations of motion, as we have previously demonstrated.

Due to the presence of additional potential terms, \cref{Hfull} is modified as
\begin{equation}
\begin{aligned}
\mathcal{H}_{eff} & = \mathcal{H}_{Kasner} + \tfrac{1}{2} e^{\tfrac{2}{\sqrt{3}} \operatorname{sgn} (\tan\beta)\left( \eta-\kappa\, J^{-1} \tan\beta \right)} C(\beta,\theta,\varphi) \\
& \qquad \qquad ~~ + \tfrac{1}{2} e^{\tfrac{1}{\sqrt{3}} \operatorname{sgn} (\tan\beta)\left( \eta-\kappa\, J^{-1} \tan\beta \right)} V(\beta,\theta,\varphi,A,E)  + \tfrac{1}{2}  W(\beta,\theta,\varphi,A,E) \,.
\end{aligned}
\end{equation}
The equations of motion acquire additional ``force'' terms compared to the system under the one-dimensional \textit{ansatz}; however, these terms are highly suppressed near the singularity. All the potential terms go exponentially to zero as $\beta\to\tfrac{\pi}{2}$, and presence of a generic electromagnetic field does not affect this behavior. This is because all components of the electromagnetic field are bounded within fixed and finite values, and thus they do not lead to any divergent contribution.

We will extend part of these results to non-Abelian gauge fields in \cref{SecB}. However, this is only true under the one-dimensional simplifying \textit{ansatz}, in which the first and the second components of the gauge fields (and momenta) are set to zero. At the moment we cannot prove the continuation result in full generality in the non-Abelian case.

\section{Extension to non-Abelian gauge fields under a one-dimensional \textit{ansatz}}
\label{SecB}

The results presented in this work for the Einstein--Maxwell--Klein--Gordon system can be extended to the Einstein--Yang--Mills--Klein--Gordon systems with $SU(2)$ and $SU(3)$ structure groups (the ones that appear in the Standard Model), under the one-dimensional \textit{ansatz}. These non-Abelian models are described by the action
\begin{equation}
\begin{array}{c}
S = \int \! d^4 x \, \sqrt{- h} \left(  R - \tfrac{1}{4}  h^{\mu\nu}  h^{\rho\sigma} F^I_{\mu\rho} F^J_{\nu\sigma}  \, \delta_{IJ} -\tfrac{1}{2} h^{\mu\nu} \partial_\mu\Phi\,\partial_\nu \Phi\right) \,, \\
\end{array}
\end{equation}
with Faraday tensor $F^I_{\mu\nu}=\partial_\mu A^I_\nu - \partial_\nu A^I_\mu + c^I_{JK} A_\mu^J A_\nu^K$. The structure constants are given by the three-dimensional Levi-Civita symbol $c^I_{JK} = \delta^{IL} \varepsilon_{LJK}$ for $SU(2)$, and, in the case of $SU(3)$, by a totally antisymmetric symbol $c^I_{JK} = \delta^{IL} f_{LJK}$, where $f_{123}=1$, $f_{147}=f_{165}=f_{246}=f_{257}=f_{345}=f_{376}={1}/{2}$, $f_{458}=f_{678}={\sqrt{3}}/{2}$, and all the others (which are not permutations of these indices) are zero. The scalar product in the internal gauge space is given by the group metric $\delta_{IJ}$, which is also used for raising and lowering internal indices.

\subsection{Homogeneous \textit{ansatz} and global constraints} In the Hamiltonian formalism, after imposing the homogeneous \textit{ansatz}, a generic Einstein--Yang--Mills--Klein--Gordon system undergoes time evolution governed by a Hamiltonian that is a linear combination of the following global constraints:
\begin{equation} \label{constraintsYM}
\begin{aligned}
\mathcal{H}[N] & = n \left(  p^{ab}p^{cd}q_{bc}\,q_{da} - \tfrac{1}{2} (p^{ab}q_{ab})^2 + q_{ab}\,q_{cd}\,\delta^{bc}\delta^{da}  - \tfrac{1}{2} ( q_{ab}\,\delta^{ab} )^2   \right.
\\ 
& \qquad \qquad\qquad \qquad\left. +\tfrac{1}{2}(p^0)^2+ \tfrac{1}{2} q_{ab}\, \delta^{IJ}E_I^a E_J^b  +   \tfrac{1}{4} \det q \, q^{ab}  q^{cd} \delta_{IJ} F^I_{ac}\, F^J_{bd}  \right) \,,
\\ \\
\mathcal{D}_i [N^i] &= n^d \left( E_I^a A^I_{b} \varepsilon_{adc}\delta^{cb}   + 2 \, p^{ab} \,q_{ac} \, \varepsilon_{bdf}\,\delta^{fc}  \right)  \,,
\\ \\
\mathcal{G}_I[A_0^I]  & = a^I_0 \left(A_a^J E^a_K c^K_{JI}\right) \,,
\end{aligned}
\end{equation}
where $\delta^{IJ}$ is the inverse group metric, $n$ and $n^a$ are defined in \cref{Lagrangemult}, and
\begin{equation}
    a_0^I=\int\! d \theta\, d\phi\, d \psi \,  \sin \theta \,  A_0^I(x) \,,
\end{equation}
are $N^2-1$ new Lagrange multipliers, corresponding to the spatial average of the scalar potential $A_0^I$, where $\operatorname{dim}SU(N)=N^2-1$ is the dimension of the gauge group. It should be noted that, unlike the Einstein--Maxwell--Klein--Gordon model where the Gauss constraint is automatically satisfied by the homogeneous ansatz (as shown in \cref{constraints}), in the case of the Einstein--Yang--Mills system, the Gauss constraint becomes a set of $N^2-1$ new proper constraints that need to be solved and gauge fixed. 

\subsection{Gauge fixing the diffeomorphism constraints}

The diffeomorphism constraints in \cref{constraintsYM} share the same functional expression as the Abelian ones (the electromagnetic contribution to the diffeomorphism constraints in \cref{constraints} can be rewritten as $E^a A_{b} \,\varepsilon_{adc}\,\delta^{cb}$). Consequently, we can use the same gauge fixing as in \cref{gaugefixing}. By solving the constraints, the following solutions are obtained:
\begin{equation}
 p^{23} = \frac  { E_I^2  A^I_3 - E_I^3  A^I_2  }{2  \left( q_2 - q_3  \right)}\,, \qquad
 p^{13}  = \frac  { E_I^3  A^I_1 - E_I^1  A^I_3}{2  \left( q_3 - q_1  \right)}\,,
\qquad
 p^{12}  = \frac  { E_I^1  A^I_2 - E_I^2  A^I_1 }{2  \left( q_1 - q_2  \right)}\,.
\end{equation}
By applying the same procedure as outlined in Section \ref{DiffeoSolution}, we derive the following on-shell Hamiltonian constraint:

\begin{equation}\label{HamConstYMuni}
\begin{aligned}
\mathcal{H}[N]  &=n \, \bigg{(} \mathcal{H}_{BIX}+\tfrac{1}{2}(p^0)^2 + \frac{q_2 q_3 (M_1)^2}{2\left(q_2-q_3\right)^2}+\frac{q_1 q_3 (M_2)^2}{2\left(q_1-q_3\right) ^2}+\frac{q_1 q_2 (M_3)^2}{2\left(q_1-q_2\right)^2}
\\ 
& \qquad ~~ + \tfrac{1}{2} \sum_{I=1}^{N^2-1}  \left(\, q_1 \!\left((E_I^1)^2 + (A^I_1)^2 \right) + q_2 \! \left((E_I^2)^2 + (A^I_2)^2 \right) +  q_3\! \left((E_I^3)^2 + (A^I_3)^2 \right) \,\right)\\ 
& \qquad ~~  + f(A A A) + g (A A A A)
  \bigg{)} \,,
\end{aligned}
\end{equation}
where, in this case, the kinetic term of the gauge fields incorporates the contribution from all the gauge components, and there are also two additional terms (cubic and quartic in the vector potential $A^I_a$) arising from the interaction of the non-Abelian gauge field with itself. These terms turn out to cancel out under the one-dimensional ansatz, so we do not find it necessary to write them out explicitly, as they have rather large expressions. The Gauss constraints, which are independent of the metric variables, remain unchanged.

\subsection{One-dimensional ansatz} As we did for the electromagnetic case, we consider a gauge field with a single spatial component:
\begin{equation}
A^I_1 =A^I_2 = 0 \,, \qquad  E_I^1 = E_I^2 = 0 \,, \qquad \forall \ I\in \{1,\dots,N^2-1\}\, .
\end{equation}
This \textit{ansatz} is well posed for the same reasons discussed at the beginning of Section \ref{uniansatz}.

The Hamiltonian constraint (\cref{HamConstYMuni}) and Gauss constraints (the last equation in \eqref{constraintsYM}) become
\begin{equation}
\begin{aligned}
\overline{\mathcal{H}}[N] & = \mathcal{H}[N]\Big|_{\substack{A^I_1 =A^I_2 = 0 \\ E_I^1 = E_I^2 = 0}} = n \left( \mathcal{H}_{BIX} +\tfrac{1}{2}(p^0)^2+  \tfrac{1}{2}\, q_3 \sum_{I=1}^{N^2-1} \!  \left( (E_I^3)^2 +  (A^I_3)^2 \right)  \right) \,,
\\ \\
\mathcal{G}_I[A_0^I]  & = a^I_0 (A_3^J\, E^3_K\, c^K_{JI}) \,.
\end{aligned}
\end{equation}
Notice that, under the one-dimensional \textit{ansatz}, the solution of the diffeomorphism constraints becomes $p^{12}=p^{23}=p^{13}=0$, similar to the one-dimensional Abelian case. Additionally, the self-interaction terms $f(AAA)$ and $f(AAAA)$ in the Hamiltonian constraint are also zero. Therefore, the Hamiltonian constraint of a one-dimensional non-Abelian system takes the same form as the Abelian one under the same \textit{ansatz}, \textit{viz.} \cref{Abelianuni}. However, the Gauss constraints still need to be solved and gauge fixed.

\subsection{Gauge fixing the Gauss constraints}
A non-Abelian model with a gauge group $SU(N)$ has $N^2-1$ nonzero Gauss constraints. However, under the one-dimensional \textit{ansatz}, not all of these constraints are independent. In the case of the groups we are interested in, namely $SU(2)$ and $SU(3)$, it is found that there are only two (out of three) linearly independent Gauss constraints for $SU(2)$, and six (out of eight) linearly independent Gauss constraints for $SU(3)$. This observation is consistent with the number of Casimir operators of these groups: $SU(2)$ has one Casimir, whereas $SU(3)$ has two. The Casimir operators represent the number of free parameters used to label the group representations, while the remaining parameters are determined by the choice of gauge for the independent Gauss constraints.

By arbitrarily selecting $\mathcal{G}_1$, $\mathcal{G}_2$ as the independent gauge generators for $SU(2)$ and $\mathcal{G}_1$, $\mathcal{G}_2$, $\mathcal{G}_4$, $\mathcal{G}_5$, $\mathcal{G}_6$, $\mathcal{G}_7$ for $SU(3)$, we can find a well-posed gauge fixing:
\begin{equation}
\begin{array}{l}
    SU(2): \ \begin{cases}
    \mathcal{G}_1 \, , \, \mathcal{G}_2 \approx 0 \, , \\
    A_3^1  \, , \, A_3^2 \approx 0 \, ,
    \end{cases} \\ \\
    SU(3): \ \begin{cases}
    \mathcal{G}_1  \, , \, \mathcal{G}_2 \, , \, \mathcal{G}_4 \, , \, \mathcal{G}_5 \, , \, \mathcal{G}_6 \, , \, \mathcal{G}_7 \approx 0 \, , \\
    A_3^1  \, , \, A_3^2 \, , \, A_3^4 \, , \, A_3^5 \, , \, A_3^6 \, , \, A_3^7 \approx 0  \, .
    \end{cases}
    \end{array}
\end{equation}
Once the Gauss constraints are solved, the conjugate momenta $E^3_I$ corresponding to the gauge-fixed components $A_3^I$ must be zero. As a result, the remaining independent gauge components are $A_3^3, E_3^3$ for $SU(2)$, and $A_3^3, A_3^8, E_3^3, E_8^3$ for $SU(3)$.

\subsection{Effective Hamiltonian constraint} After solving the Gauss constraints, the only remaining constraint is the Hamiltonian one:
\begin{equation}
    \begin{array}{l}
  SU(2): \ \overline{\mathcal{H}}[N]=n \left( \mathcal{H}_{BIX} +\tfrac{1}{2}(p^0)^2 +  \tfrac{1}{2}\, q_3  \left( (E_3^3)^2 +  (A^3_3)^2 \right)  \right) \,, \\ \\
   SU(3): \ \overline{\mathcal{H}}[N]=n \left( \mathcal{H}_{BIX} +\tfrac{1}{2}(p^0)^2 +  \tfrac{1}{2}\, q_3  \left( (E_3^3)^2+E_8^3)^2 +  (A^3_3)^2 +(A^8_3)^2 \right)  \right) \,.
    \end{array}
\end{equation}
As discussed in Section \ref{uniansatz}, we can identify a conserved quantity, which is a first-class quantity with respect to the Hamiltonian constraint. In the case of $SU(2)$, this conserved quantity corresponds again to a one-dimensional harmonic oscillator,
\begin{equation}
    H_{HO}^{1D}=(E_3^3)^2 + (A^3_3)^2 \,,
\end{equation}
(compare this to the Abelian case, \cref{ho1d}), while for $SU(3)$ the conserved quantity is
\begin{equation}
    H_{HO}^{2D}=(E_3^3)^2+E_8^3)^2 +  (A^3_3)^2 +(A^8_3)^2 \,.
\end{equation}
$H_{HO}^{1D}$ and $H_{HO}^{2D}$ are constants of motion, and we can set them to a positive constant $\varepsilon$ for both gauge groups without loss of generality. This final step allows us to describe the dynamics of both $SU(2)$ and $SU(3)$ gauge systems with the same effective Hamiltonian:
\begin{equation}
    \mathcal{H}_{eff}[N]  = n \left( \mathcal{H}_{BIX} +\tfrac{1}{2}(p^0)^2 +  {\tfrac{1}{2}} \, q_3  \, \varepsilon  \right) \,.
\end{equation}
Since this effective Hamiltonian is identical to that of the one-dimensional Abelian case (see \cref{Hconstraint}), the continuation result proven in Section \ref{uniansatz} also applies to the Einstein--Yang--Mills--Klein--Gordon systems with $SU(2)$ and $SU(3)$ as structure groups under the one-dimensional \textit{ansatz}.

\section{Conclusions}
\label{Sec8}

In Ref.~\cite{ThroughTheBigBang}, we conjectured that it is possible to continue Einstein's \textit{classical} equations through the big bang singularity into another universe with an opposite time direction and spatial orientation, which preserves all the information about the state of the universe on the other side of the singularity (although it might become irretrievably scrambled in the process due to a chaotic phase of the dynamics). This is intimately related to far-reaching issues such as black hole unitarity and the nature of the big bang.

This conjecture was proven in simplified cases, including homogeneous cosmologies \cite{ThroughTheBigBang}, inflationary models \cite{Mercati_2019,Sloan:2019wrz}, and the Schwarzschild-scalar system \cite{Mercati_2022}. Our approach is to gradually increase the complexity of the models under consideration, test the validity of the conjecture, and gain insight into the behavior of physical fields across the singularity. As mentioned in the previous paper \cite{Mercati_2019}, the next natural step in this process is to determine if the predicted reversal of orientation at the singularity can be physically measurable. In other words, can the inhabitants of the universe determine, through an experiment, which side of the big bang singularity they live in?

To answer this question, three ingredients are necessary. First, we need to understand what happens to the orientation of space defined by the vielbein/frame fields. It has been established that these fields undergo a sign change at the singularity in the original paper \cite{ThroughTheBigBang}. Second, we must establish the behavior of vector (gauge) fields and fermions. If all of these fields undergo a ``flipping'' transition at the singularity, it might cancel out the orientation reversal effect of the vielbeins, making it unobservable. Finally, we need to investigate what happens to experimentally realizable processes, such as beta decays. Ultimately, the crucial factor is whether the parity-breaking vertices of the Standard Model remain unchanged across the singularity when considering their dependence on the spacetime vielbeins.

In the present paper, we conducted a detailed analysis of gauge fields. We first determined that the continuation result remains unchanged in the presence of Abelian gauge fields (in general) and non-Abelian gauge fields (under the simplifying assumption of the one-dimensional ansatz). Additionally, we established that the behavior of the gauge fields near the singularity is straightforward: their values freeze, with zero time derivatives at the exact instant of the singularity, and they evolve through it without flipping their orientation. The next logical step is to analyze fermion fields, which will allow us to determine the fate of the parity-breaking vertices of the Standard Model. Another interesting extension of this work would be to relax the one-dimensional ansatz for non-Abelian gauge fields, although this step has not been feasible thus far. This paper provides compelling evidence that the general case, beyond the one-dimensional \textit{ansatz}, does not affect the continuation outcome nor the conclusion that gauge fields do not ``flip'' at the singularity. However, there is still some uncertainty around this matter, and further research is needed for confirmation.

Our work offers a somewhat complementary perspective on singularities, compared for example to works like~\cite{Wald:1980jn, Ishibashi:1999vw,Ishibashi:2003jd} and similar ones, which assumed a fixed background spacetime with a singularity, and studied whether, and under which conditions, the propagation of matter fields on such background can be deterministic. Our conjecture, namely that when expressed in the appropriate variables the full gravity+matter dynamics remains classically deterministic, relies crucially on taking into account the backreaction of matter fields on the geometry. However, as shown in the present paper, near the singularity there is a form of decoupling between the matter and gravitational degrees of freedom, which makes the backreaction increasingly irrelevant as the singularity is approached, and therefore if the deterministic evolution of the gravity degrees of freedom can be proven, the matter ones might be treated in a similar fashion to the approach of papers like \cite{Wald:1980jn, Ishibashi:1999vw,Ishibashi:2003jd}, at least asymptotically. This is an interesting starting point for further explorations.

A final note on quantum gravity: one of the most intriguing aspects of our approach, in our opinion, is that the singularity resolution mechanism discovered in \cite{ThroughTheBigBang} is entirely classical, and does not rely on quantum effects. Moreover, this mechanism is entirely infrared, involving the homogeneous (\textit{i.e.}, the ``most infrared'')  degrees of freedom. This suggests that, contrary to expectations, the singularity might be a relatively benign phenomenon. It does not necessarily involve the excitation of deep ultraviolet degrees of freedom, and might, in fact, be well within the domain of validity of perturbative quantum gravity. This is an exciting prospect that motivates interest in studying the physics of quantum fluctuations on our singularity-resolving background spacetime. However, such interest would disappear entirely if it turned out that the introduction of other Standard Model fields destroys the original continuation result, as we know that such fields exist and have to be taken into account.
We therefore postpone the investigation of quantum effects to a time when the result for the classical model, with all the matter fields of the Standard Model, has been proven and rests on firm grounds.

\section*{Acknowledgments}

This work has been supported by grant PID2023-148373NB-I00, funded by\\ MCIN/AEI/10.13039/501100011033/FEDER--EU, and by the Q-CAYLE Project, funded by the Regional Government of Castilla y León (Junta de Castilla y León) and the Ministry of Science and Innovation (MCIN) through the European Union's NextGenerationEU funds (PRTR C17.I1).
F.M.~acknowledges support from the Agencia Estatal de Investigación (Spain) under grant CNS2023-143760.

\small
\bibliographystyle{utphys}
\bibliography{Through_Gauge_Bib.bib}

\end{document}